\documentclass[preprint2]{aastex}
\usepackage{natbib}
\begin{document}

\title{\large \rm THE SCALE OF REDDENING FOR CLASSICAL CEPHEID VARIABLES}
\author{\sc \small D.~G. Turner}
\affil{\footnotesize Saint Mary's University, Halifax, NS, B3H 3C3, Canada.}
\email{turner@ap.smu.ca}

\begin{abstract}
Field reddenings are summarized for 68 Cepheids from published studies and updated results presented here. The compilation forms the basis for a comparison with other published reddening scales of Cepheids, including those established from reddening-independent indices, photometry on the Lick six-color system, Str\"{o}mgren system, Walraven system, Washington system, Cape {\it BVI} system, DDO system, and Geneva system, IRSB studies, and Cepheid spectroscopy, both old and new. Reddenings tied to period-color relations are the least reliable, as expected, while photometric color excesses vary in precision, their accuracy depending on the methodology and calibration sample. The tests provide insights into the accuracy and precision of published Cepheid reddening scales, and lead to a new system of standardized reddenings comprising a sample of 198 variables with an average uncertainty of $\pm0.028$ in E$_{B-V}$, the precision being less than $\pm0.01$ for many. The collected color excesses are used to map the dispersion in intrinsic colors as a function of pulsation period, the results contradicting current perceptions about the period dependence of dispersion in Cepheid effective temperatures.
\end{abstract}

\section{{\rm \footnotesize INTRODUCTION}}

Classical Cepheids are pulsating yellow supergiants that make ideal standard candles for distance estimates because of the fairly tight relationship that exists between their luminosities and periods of pulsation --- the period-luminosity relation, also referred to as the Leavitt law when expressed as absolute magnitude versus period. The general properties of Cepheids consequently make them of considerable importance for the reliability of the distance scale. That, in turn, has inspired a quest, still ongoing, to delineate the Cepheid instability strip as accurately as possible \citep[see][]{tu12b}. As noted by \citet*{te11}, an empirical description of the instability strip has been difficult to establish because of complications arising from Galactic reddening and extinction; extinction alone represents an overlooked obstacle because of the prescriptive fashion by which it is often applied \citep[see][]{tu14}. But the creation of an accurate scale of Cepheid reddenings (color excesses) represents an important first step.

Historically the color excesses of Cepheids have been established using field reddenings derived from photometric and spectroscopic observations of their nearby early-type companions, using photometric or spectrophotometric observations of the variables in calibrated reddening-independent systems, from comparison of their cyclical spectroscopic and photometric variations with derived intrinsic colors for supergiants of comparable spectral types, and from simple period-color relations tied to a few standard Cepheids calibrated by the previous methods. Each method is fraught with difficulties arising from uncertainties in the methods of establishing intrinsic colors for luminous stars of all types, as well as dereddening them, given that few are close enough in the Galactic plane to be unaffected by the contaminating effects of nearby dust clouds.

The last technique, however, suffers from a more fundamental problem: the intrinsic width of the instability strip in effective temperature, combined with the fact that stars of identical mean radius (or pulsation period) lie diagonally across the strip over a range of mean effective temperatures, implies that stars of identical period can have quite different intrinsic colors, in conflict with the basic premise of a period-color relation. Coincidentally, many well-studied bright Cepheids have large light amplitudes, and such variables are usually found in the more constricted region of effective temperature lying close to and slightly warmer than the center line of the instability strip \citep*[see][]{te06a}. Adoption of period-color relations to deredden Cepheids can therefore yield reasonable color excesses in some cases, namely Cepheids of large amplitude, although the scatter inherent to the methodology must propagate into uncertainties for individual stars.

The accuracy and precision of reddenings for early-type stars have gradually improved over the past 60 years through developments in dereddening techniques and more reliable intrinsic colors. In parallel to such advances, \citet{ko08} have increased the precision of spectroscopic parameters for Cepheids and yellow supergiants through use of as many temperature-sensitive and gravity-sensitive line ratios as permitted by the dimensions of their observational spectral windows. Their technique eliminates any dependence on adopted reddening relations by matching stellar atmosphere models to Cepheid spectra over their cycles to infer effective temperatures, and hence intrinsic colors, for comparison with observed colors. Older versions of Cepheid dereddening procedures from spectra were generally of much lower precision.

An oft-cited source of Cepheid reddenings was established 25 years ago by \citet{fe90a} and used to study the Cepheid instability strip \citep{fe90b}. The intrinsic relations adopted for that compilation were calibrated using a variety of different procedures noted above \citep{fe87}, resulting in zero-point offsets with respect to other reddening calibrations of that era. \citet{tu95}, for example, noted the existence of systematic offsets in reddening for some Cepheids, notably long period variables.

The importance of accurate Cepheid reddenings has been pointed out previously \citep{tb02,tu10,tu12b}. A Cepheid's mean radius is closely related to its pulsation period according to independent studies using the Baade-Wesselink method and its variants \citep{ls93,gi98,ar00,tb02,te10}, while its average effective temperature can be established fairly reliably from its unreddened broad band {\it B--V} color \citep{gr92,tb02}, as also argued by \citet{fc99}. Combining the two values then yields a Cepheid's mean luminosity according to the standard relationship, $\langle L \rangle=4\pi \langle R \rangle ^2 \sigma \langle T_{\rm eff} \rangle ^4$, where angular brackets denote averages over the pulsation cycle. The Cepheid period-luminosity relation can therefore be constructed from basic information on the pulsation period and reddening of Cepheids in combination with observed {\it B--V} colors \citep{tu12b}. The distribution of Cepheids in period-color space can also be established in similar fashion \citep[see][]{tu01,tu10} to provide an empirical description of the Cepheid instability strip.

The gradual growth in field reddenings for Galactic Cepheids \citep[e.g.,][]{lc07} has made it possible to make detailed comparisons of existing reddening compilations in order to examine the accuracy and precision of each. The present study was therefore initiated in order to compile all available field reddenings for Cepheids, to redo any where reanalysis was warranted, and to use the resulting system of reddenings to examine in detail a selection of existing reddening compilations, including those often cited for the extragalactic Cepheid calibration. 

\section{{\rm \footnotesize THE REDDENING SYSTEM}}

The reddening system established here is based upon the growing collection of space reddenings for Cepheids that has developed over the past thirty-five years, mostly as by-products of photometric studies of open clusters and associations containing Cepheids as potential calibrators for the period-luminosity relation. The collection is not complete (several studies are still in progress), but is large enough for the present purpose.

\setcounter{table}{0}
\begin{table}[!t]
\caption[]{\small{Intrinsic {\it UBV} Colors for Zero-Age Zero-Rotation Dwarfs.}}
\label{tab1}
\centering
\tiny
\begin{tabular*}{0.47\textwidth}{@{\extracolsep{-3.7mm}}cccccccccc}
\hline \hline \noalign{\smallskip}
({\it B--V})$_0$ &({\it U--B})$_0$ &({\it B--V})$_0$ &({\it U--B})$_0$ &({\it B--V})$_0$ &({\it U--B})$_0$ &({\it B--V})$_0$ &({\it U--B})$_0$ &({\it B--V})$_0$ &({\it U--B})$_0$ \\
\noalign{\smallskip} \hline \noalign{\smallskip}
$-0.33$ &$-1.205$ &$+0.05$ &$+0.083$ &$+0.43$ &$-0.020$ &$+0.81$ &$+0.425$ &$+1.19$ &$+1.086$ \\
$-0.32$ &$-1.171$ &$+0.06$ &$+0.091$ &$+0.44$ &$-0.021$ &$+0.82$ &$+0.444$ &$+1.20$ &$+1.097$ \\
$-0.31$ &$-1.133$ &$+0.07$ &$+0.097$ &$+0.45$ &$-0.022$ &$+0.83$ &$+0.464$ &$+1.21$ &$+1.109$ \\
$-0.30$ &$-1.089$ &$+0.08$ &$+0.101$ &$+0.46$ &$-0.022$ &$+0.84$ &$+0.485$ &$+1.22$ &$+1.120$ \\
$-0.29$ &$-1.047$ &$+0.09$ &$+0.104$ &$+0.47$ &$-0.020$ &$+0.85$ &$+0.506$ &$+1.23$ &$+1.132$ \\
$-0.28$ &$-1.008$ &$+0.10$ &$+0.105$ &$+0.48$ &$-0.018$ &$+0.86$ &$+0.527$ &$+1.24$ &$+1.143$ \\
$-0.27$ &$-0.969$ &$+0.11$ &$+0.105$ &$+0.49$ &$-0.014$ &$+0.87$ &$+0.549$ &$+1.25$ &$+1.154$ \\
$-0.26$ &$-0.932$ &$+0.12$ &$+0.104$ &$+0.50$ &$-0.008$ &$+0.88$ &$+0.570$ &$+1.26$ &$+1.166$ \\
$-0.25$ &$-0.895$ &$+0.13$ &$+0.103$ &$+0.51$ &$-0.002$ &$+0.89$ &$+0.592$ &$+1.27$ &$+1.177$ \\
$-0.24$ &$-0.859$ &$+0.14$ &$+0.100$ &$+0.52$ &$+0.007$ &$+0.90$ &$+0.615$ &$+1.28$ &$+1.189$ \\
$-0.23$ &$-0.823$ &$+0.15$ &$+0.098$ &$+0.53$ &$+0.016$ &$+0.91$ &$+0.637$ &$+1.29$ &$+1.200$ \\
$-0.22$ &$-0.787$ &$+0.16$ &$+0.095$ &$+0.54$ &$+0.023$ &$+0.92$ &$+0.660$ &$+1.30$ &$+1.211$ \\
$-0.21$ &$-0.751$ &$+0.17$ &$+0.091$ &$+0.55$ &$+0.033$ &$+0.93$ &$+0.682$ &$+1.31$ &$+1.221$ \\
$-0.20$ &$-0.715$ &$+0.18$ &$+0.088$ &$+0.56$ &$+0.044$ &$+0.94$ &$+0.704$ &$+1.32$ &$+1.231$ \\
$-0.19$ &$-0.678$ &$+0.19$ &$+0.084$ &$+0.57$ &$+0.055$ &$+0.95$ &$+0.725$ &$+1.33$ &$+1.240$ \\
$-0.18$ &$-0.641$ &$+0.20$ &$+0.080$ &$+0.58$ &$+0.066$ &$+0.96$ &$+0.745$ &$+1.34$ &$+1.248$ \\
$-0.17$ &$-0.604$ &$+0.21$ &$+0.076$ &$+0.59$ &$+0.078$ &$+0.97$ &$+0.766$ &$+1.35$ &$+1.255$ \\
$-0.16$ &$-0.566$ &$+0.22$ &$+0.072$ &$+0.60$ &$+0.091$ &$+0.98$ &$+0.786$ &$+1.36$ &$+1.261$ \\
$-0.15$ &$-0.528$ &$+0.23$ &$+0.068$ &$+0.61$ &$+0.104$ &$+0.99$ &$+0.805$ &$+1.37$ &$+1.266$ \\
$-0.14$ &$-0.489$ &$+0.24$ &$+0.064$ &$+0.62$ &$+0.118$ &$+1.00$ &$+0.824$ &$+1.38$ &$+1.269$ \\
$-0.13$ &$-0.451$ &$+0.25$ &$+0.059$ &$+0.63$ &$+0.132$ &$+1.01$ &$+0.842$ &$+1.39$ &$+1.271$ \\
$-0.12$ &$-0.412$ &$+0.26$ &$+0.055$ &$+0.64$ &$+0.146$ &$+1.02$ &$+0.860$ &$+1.40$ &$+1.271$ \\
$-0.11$ &$-0.373$ &$+0.27$ &$+0.050$ &$+0.65$ &$+0.160$ &$+1.03$ &$+0.877$ &$+1.41$ &$+1.270$ \\
$-0.10$ &$-0.334$ &$+0.28$ &$+0.046$ &$+0.66$ &$+0.175$ &$+1.04$ &$+0.894$ &$+1.42$ &$+1.266$ \\
$-0.09$ &$-0.295$ &$+0.29$ &$+0.041$ &$+0.67$ &$+0.190$ &$+1.05$ &$+0.910$ &$+1.43$ &$+1.261$ \\
$-0.08$ &$-0.257$ &$+0.30$ &$+0.037$ &$+0.68$ &$+0.205$ &$+1.06$ &$+0.925$ &$+1.44$ &$+1.254$ \\
$-0.07$ &$-0.220$ &$+0.31$ &$+0.032$ &$+0.69$ &$+0.221$ &$+1.07$ &$+0.940$ &$+1.45$ &$+1.246$ \\
$-0.06$ &$-0.181$ &$+0.32$ &$+0.027$ &$+0.70$ &$+0.236$ &$+1.08$ &$+0.954$ &$+1.46$ &$+1.236$ \\
$-0.05$ &$-0.147$ &$+0.33$ &$+0.022$ &$+0.71$ &$+0.252$ &$+1.09$ &$+0.967$ &$+1.47$ &$+1.224$ \\
$-0.04$ &$-0.113$ &$+0.34$ &$+0.017$ &$+0.72$ &$+0.268$ &$+1.10$ &$+0.980$ &$+1.48$ &$+1.213$ \\
$-0.03$ &$-0.081$ &$+0.35$ &$+0.013$ &$+0.73$ &$+0.284$ &$+1.11$ &$+0.993$ &$+1.49$ &$+1.203$ \\
$-0.02$ &$-0.051$ &$+0.36$ &$+0.008$ &$+0.74$ &$+0.301$ &$+1.12$ &$+1.006$ &$+1.50$ &$+1.191$ \\
$-0.01$ &$-0.022$ &$+0.37$ &$+0.003$ &$+0.75$ &$+0.317$ &$+1.13$ &$+1.018$ &$+1.51$ &$+1.178$ \\
$+0.00$ &$+0.000$ &$+0.38$ &$-0.002$ &$+0.76$ &$+0.334$ &$+1.14$ &$+1.029$ &$+1.52$ &$+1.166$ \\
$+0.01$ &$+0.026$ &$+0.39$ &$-0.006$ &$+0.77$ &$+0.352$ &$+1.15$ &$+1.041$ \\
$+0.02$ &$+0.044$ &$+0.40$ &$-0.010$ &$+0.78$ &$+0.369$ &$+1.16$ &$+1.052$ \\
$+0.03$ &$+0.060$ &$+0.41$ &$-0.015$ &$+0.79$ &$+0.388$ &$+1.17$ &$+1.064$ \\
$+0.04$ &$+0.073$ &$+0.42$ &$-0.018$ &$+0.80$ &$+0.406$ &$+1.18$ &$+1.075$ \\
\noalign{\smallskip} \hline
\end{tabular*}
\end{table}

Reddenings were rederived for a selection of Cepheids studied previously, namely those of \citet{fm76}, \citet{tu80a}, \citet{te05}, and \citet{tu13}. The analyses made use of Str\"{o}mgren system data transformed to the {\it UBV} system as described by \citet{tu90} --- see, as an example, \citet{tu12a}, --- existing data, and {\it BV} data supplemented by new CCD observations from the AAVSO (American Association of Variable Star Observers) Photometric All-Sky Survey (APASS).

Thirty years ago it was common practice to adopt intrinsic {\it UBV} colors for A-type dwarfs from compilations such as those of \citet{jo66} and \citet{fi70}. Later studies \citep{cs85,ps87} revealed that such colors are likely to be biased by a mix of stars having different rotational velocities and projection angles. True zero-age zero-rotation main-sequence (ZAZRMS) stars of spectral types A and F have colors that differ systematically from those of \citet{jo66} and \citet{fi70} according to \citet{ps87}, although the standardization for the latter study was mistakenly tied to A0 stars with colors affected by a zero-point offset. True ZAZRMS colors for dwarfs were therefore recompiled by smoothing the colors for OB and KM stars from \citet{jo66} and \citet{fi70}, renormalizing the \citet{ps87} Johnson and Str\"{o}mgren system intrinsic color predictions for non-rotating AF dwarfs to A0 stars with $[(B-V)_0,(U-B)_0]=[0.00,0.00]$, and adding the colors for unreddened Pleiades GK dwarfs (roughly solar metallicity) from the study by \citet{tu77b}.

\begin{figure}[ht]
\begin{center}
\includegraphics[width=0.45\textwidth]{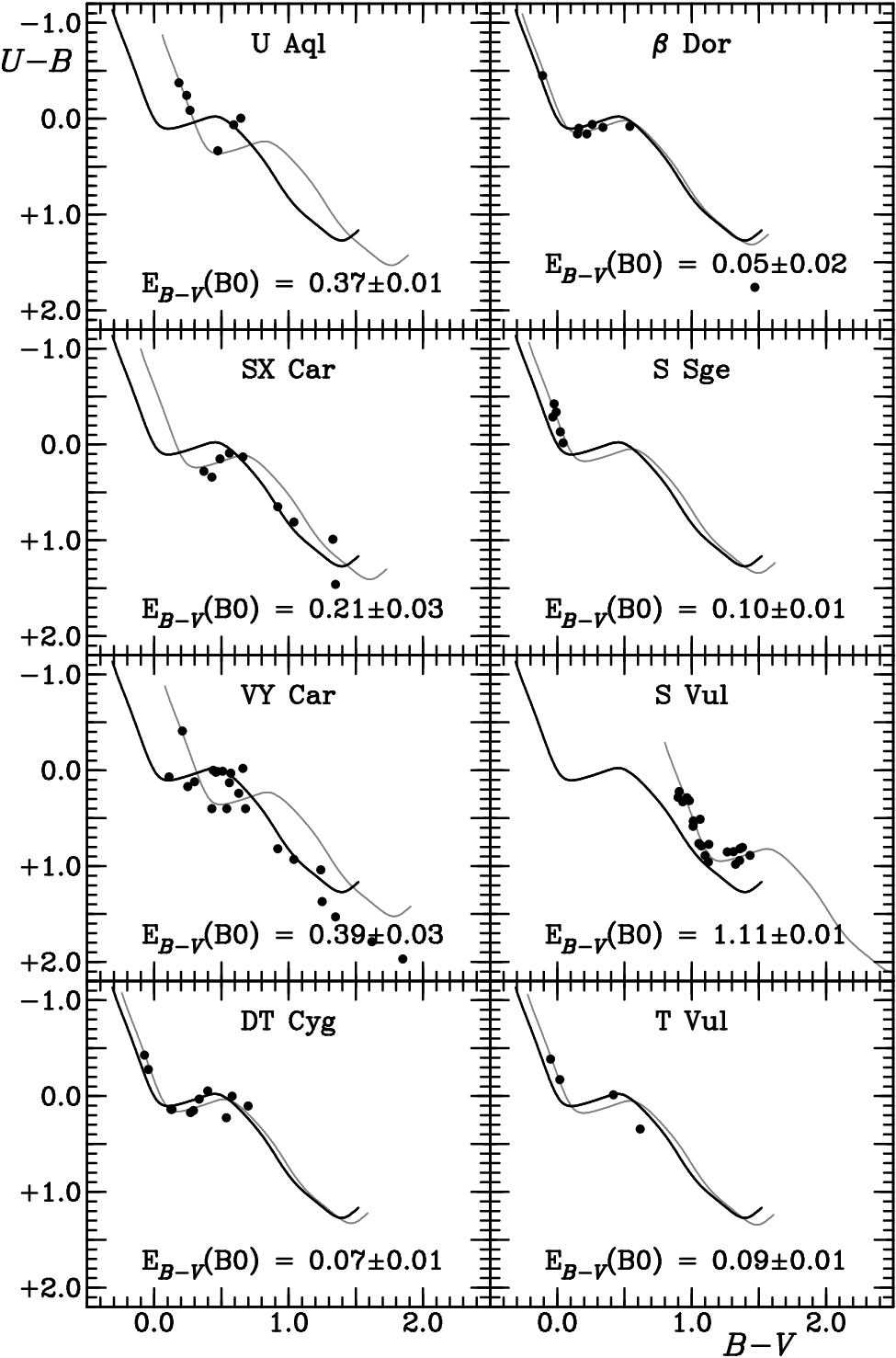}
\end{center}
\caption{\small{Existing and transformed {\it UBV} data for stars surrounding eight Cepheids. The intrinsic ZAZRMS color-color relation is shown (in black) in each case, along with the same relation reddened by the average B0-star color excess (indicated in each case and depicted by the gray lines) for stars associated with each Cepheid.}}
\label{fig1}
\end{figure}

The resulting {\it UBV} colors were shown previously to match the observed colors of OB and AF stars in uniformly-reddened open clusters \citep{tu96}. Prior to that they were tested and confirmed using observed colors for large samples of BAF-type stars from the {\it Photoelectric Catalogue} of \citet{bl70}. Only stars considered to have {\it UBV} photometry properly tied to the Johnson system were used for testing. The resulting intrinsic ZAZRMS relation is shown in Figs.~2~\&~3 of \citet{tu96} and here in Fig.~\ref{fig1}, but for the benefit of potential users is also tabulated in Table~\ref{tab1}. Fig.~\ref{fig1} depicts existing and transformed {\it UBV} data used to establish new space reddenings for the eight Cepheids noted above. 

An overlooked feature of interstellar reddening is that no single relationship accurately describes it over all regions of the Galaxy \citep*{wa61,wa62,ma90,tu89,tu94,zg12,zg13,te14}. Regional variations in reddening law display an inherent dependence upon direction viewed through the Galaxy, so the reddenings derived in the color-color diagrams of Fig.~\ref{fig1} were inferred using reddening laws, {\it i.e.,} $E_{U-B}/E_{B-V}$, appropriate for each field. The corresponding color excesses for the Cepheids were also adjusted from their averaged equivalent B0-star reddenings using the transformation of \citet{fe63}. Such corrections are fairly straightforward given that the mean colors of each Cepheid are known, but do depend upon the accuracy of the adopted \citet{fe63} relationship.

Another misconception is that reddening increases linearly with distance \citep[e.g.,][]{lc07}, whereas in reality reddening increases discretely whenever the line of sight crosses an interstellar (dust) cloud causing extinction \citep[see][]{te11}. In some directions, such as a few degrees away from the Galactic equator, all reddened stars may be affected by light passing through only one relatively nearby dust cloud, no matter how distant they lie beyond the cloud. That makes the derivation of space reddening for many Cepheids relatively uncomplicated. They need not belong to a cluster or association, as long as there are some early-type stars in the same line of sight.

A problem arose with the reddening for the Cepheids QZ~Nor and V340~Nor, associated with the open cluster NGC~6067. \citet{ma13a} adopted reddenings of E$_{J-H} = 0.13$ for the cluster and Cepheids based upon a spectroscopic reddening of E$_{B-V}$(B0) = $0.35 \pm 0.04$ tied to spectra and {\it UBV} photometry by \citet*{th62} for NGC~6067. A re-examination of the {\it UBV} colors for cluster stars from \citet{th62} and \citet{pd84} in conjunction with a likely reddening law for the field \citep[see][]{tu76c,te14} revealed that the cluster displays differential reddening. Color excesses for cluster stars vary from E$_{B-V} = 0.30$ to E$_{B-V} = 0.48$, with most stars reddened by 0.36--0.42. The E$_{J-H}$ reddening adopted by \citet{ma13a} corresponds to E$_{B-V} = 0.42$ with the relations of \citet{tu11}, in contrast to the mean spectroscopic reddening. Since V340~Nor lies in the nuclear region of NGC~6067 while QZ~Nor is a coronal member, the field reddening for both Cepheids must be examined separately.

The available data for stars lying within 2 arcmin of each Cepheid were therefore reanalyzed using the methodology described previously \citep[see also][]{tu96}. The observations were primarily the {\it UBV} data of \citet{th62} and \citet{pd84}, {\it BV} data from APASS, and {\it JHK}$_s$ data \citep{cu03} from the 2MASS survey \citep{sk06}. Stars near V340~Nor displayed a reddening of E$_{B-V} = 0.39 \pm0.02$ (transformed from E$_{J-H} = 0.115 \pm 0.005$), which was adopted for the Cepheid space reddening given in Table~\ref{tab2}. There were fewer stars available to test the space reddening near QZ~Nor, but they did appear to be consistent with a color excess identical to that for V340~Nor. That reddening was therefore adopted for the space reddening  of QZ~Nor in Table~\ref{tab2}.

\setcounter{table}{1}
\begin{table*}[htp]
\caption[]{\small{Space Reddenings for Cepheids.}}
\label{tab2}
\begin{center}
\small
\begin{tabular*}{1.00\textwidth}{@{\extracolsep{+0.1mm}}llclcllcl}
\hline \hline
Cepheid &Field &E$_{B-V}$ &Source & &Cepheid &Field &E$_{B-V}$ &Source \\
\hline
T~Ant &Anon~Ant~OB &$0.30\pm0.01$ &24 & &AC~Mon &Field &$0.55\pm0.01$ &34 \\
U~Aql &Field &$0.35\pm0.01$ &2, 44 & &CV~Mon &van~den~Bergh~1 &$0.75\pm0.02$ &22 \\
FF~Aql &Field &$0.25\pm0.01$ &28, 41 & &S~Nor &NGC~6087 &$0.17\pm0.01$ &12 \\
RT~Aur &Field &$0.06\pm0.03$ &28 & &TW~Nor &Lyng\"{a}~6 &$1.29\pm0.10$ &35 \\
EW~Aur &Field &$0.58\pm0.03$ &34 & &QZ~Nor &NGC~6067 &$0.39\pm0.02$ &42, 44 \\
OX~Cam &Tombaugh~5 &$0.75\pm0.07$ &29 & &V340~Nor &NGC~6067 &$0.39\pm0.02$ &42, 44 \\
XZ~CMa &Tombaugh~1 &$0.31\pm0.01$ &9 & &UY~Per &King~4 &$0.89\pm0.05$ &33 \\
YZ~CMa &Field &$0.56\pm0.03$ &34 & &HZ~Per &Field &$1.36\pm0.04$ &34 \\
CN~CMa &Field &$0.63\pm0.02$ &34 & &OT~Per &Field &$1.39\pm0.08$ &34 \\
$\ell$~Car &Field &$0.17\pm0.02$ &28 & &AQ~Pup &Turner~14 &$0.47\pm0.07$ &39 \\
U~Car &Anon~Car~OB &$0.30\pm0.03$ &14 & &BD~Pup &Field &$0.67\pm0.02$ &34 \\
SX~Car &Ruprecht~91 &$0.20\pm0.03$ &25, 44 & &BE~Pup &Field &$0.64\pm0.02$ &34 \\
VY~Car &Ruprecht~91 &$0.36\pm0.03$ &25, 44 & &LR~Pup &Field &$0.42\pm0.01$ &34 \\
WZ~Car &Collinder~236 &$0.27\pm0.01$ &32 & &V620~Pup &Turner~13 &$0.64\pm0.02$ &39 \\
GT~Car &Teutsch~106 &$0.66\pm0.02$ &17 & &S~Sge &Field &$0.09\pm0.01$ &2, 44 \\
SU~Cas &Alessi~90 &$0.33\pm0.02$ &38 & &GY~Sge &Sge~OB1 &$1.15\pm0.02$ &6 \\
CE~Cas &NGC~7790 &$0.49\pm0.05$ &43 & &U~Sgr &M25 &$0.49\pm0.03$ &43 \\ 
CF~Cas &NGC~7790 &$0.49\pm0.05$ &43 & &W~Sgr &Field &$0.12\pm0.01$ &28 \\
CG~Cas &Berkeley~58 &$0.64\pm0.02$ &30 & &X~Sgr &Field &$0.19\pm0.01$ &28 \\
DL~Cas &NGC~129 &$0.51\pm0.01$ &16 & &Y~Sgr &Field &$0.22\pm0.01$ &28 \\
FO~Cas &Field &$0.76\pm0.05$ &34 & &VY~Sgr &Field &$1.24\pm0.04$ &34\\
IO~Cas &Field &$0.59\pm0.02$ &34 & &WZ~Sgr &Turner~2 &$0.56\pm0.01$ &17  \\
V~Cen &NGC~5662b &$0.28\pm0.01$ &8 & &AY~Sgr &Field &$0.94\pm0.02$ &34 \\
V810~Cen &Stock~14 &$0.25\pm0.01$ &7 & &BB~Sgr &Collinder~394 &$0.30\pm0.02$ &10, 11 \\
$\delta$~Cep &Cep~OB6 &$0.07\pm0.01$ &23, 36 & &V1882~Sgr &Field &$0.64\pm0.01$ &34 \\
X~Cyg &Ruprecht~173/5 &$0.25\pm0.02$ &21 & &RU~Sct &Trumpler~35 &$0.95\pm0.02$ &4 \\
SU~Cyg &Turner~9 &$0.15\pm0.01$ &19, 20 & &V367~Sct &NGC~6649 &$1.27\pm0.02$ &5 \\
DT~Cyg &Field &$0.07\pm0.01$ &2, 44 & &SZ~Tau &NGC~1647 &$0.29\pm0.01$ &9, 15 \\
V1726~Cyg &Platais~1 &$0.33\pm0.02$ &18, 27 & &SW~Vel &Vel~OB1c &$0.35\pm0.02$ &17 \\
$\beta$~Dor &Field &$0.05\pm0.02$ &3, 28, 44 & &CS~Vel &Ruprecht~79 &$0.75\pm0.01$ &13 \\
$\zeta$~Gem &Field &$0.02\pm0.01$ &28, 37 & &S~Vul &Turner~1 &$1.01\pm0.01$ &44 \\
HD~18391 &TKMLM~1 &$1.02\pm0.09$ &31 & &T~Vul &Field &$0.11\pm0.02$ &2, 28, 44 \\
T~Mon &Mon~OB~2 &$0.23\pm0.04$ &1 & &SV~Vul &Vul~OB1 &$0.45\pm0.01$ &9 \\
UY~Mon &Field &$0.11\pm0.01$ &34 & &$\alpha$~UMi &Harrington~1 &$0.02\pm0.01$ &26, 40 \\
\hline
\end{tabular*}
\footnotesize
\end{center}
Source: (1) \citet{tu76b}, (2) \citet{fm76}, (3) \citet{tu80a}, (4) \citet{tu80b}, (5) \citet{tu81}, (6) \citet{fb82}, (7) \citet{tu82a}, (8) \citet{tu82b}, (9) \citet{tu83}, (10) \citet{tu84}, (11) \citet{te85}, (12) \citet{tu86}, (13) \citet{wa87}, (14) \citet{tu88}, (15) \citet{tu92}, (16) \citet{te92}, (17) \citet{te93}, (18) \citet{te94}, (19) \citet{te97}, (20) \citet{te98a}, (21) \citet{tu98}, (22) \citet{te98b}, (23) \citet{bn02}, (24) \citet{tb03}, (25) \citet{te05}, (26) \citet{tu05}, (27) \citet{te06b}, (28) \citet{bn07}, (29) \citet{ma08}, (30) \citet{te08}, (31) \citet{te09a}, (32) \citet{te09b}, (33) \citet{te10}, (34) \citet{te11}, (35) \citet{ma11}, (36) \citet{ma12a}, (37) \citet{ma12b}, (38) \citet{te12a}, (39) \citet{te12b}, (40) \citet{te13a}, (41) \citet{te13b}, (42) \citet{ma13a}, (43) \citet{ma13b}, (44) This paper.
\end{table*}

The sample of 68 Cepheids of known space reddening is summarized in Table~\ref{tab2}, where the sources used to establish the color excesses are indicated in the Table notes. The reddening of each Cepheid is based upon studies of the surrounding field in which the exact form of the reddening law in the field was (usually) established beforehand. The space reddening sample is clearly quite extensive, although a few of the Cepheids are of uncertain use given questions about their population types \citep[see][]{te11}.

\section{{\rm \footnotesize TESTING THE SAMPLE}}
\begin{figure}[h]
\begin{center}
\includegraphics[width=0.45\textwidth]{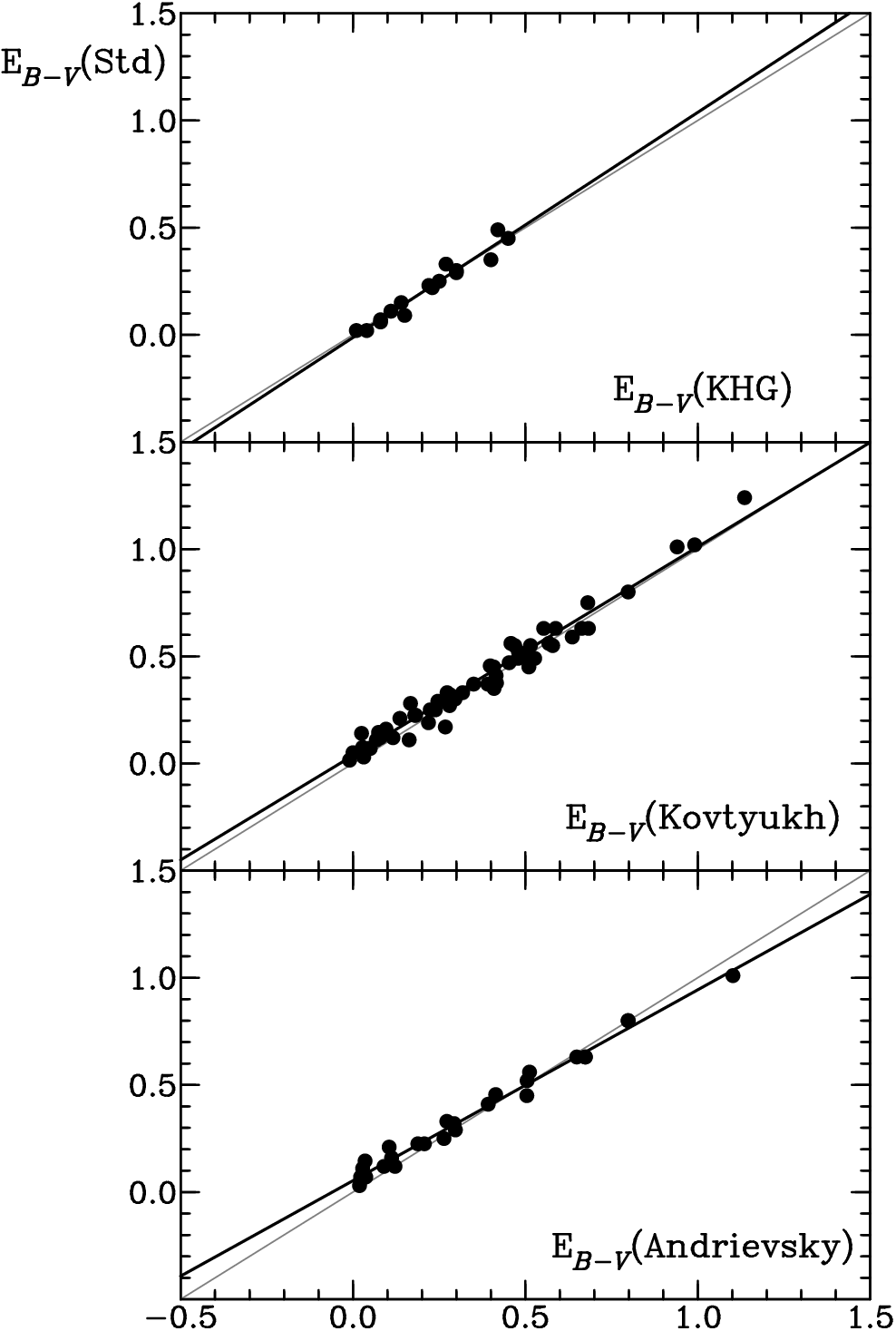}
\end{center}
\caption{\small{A comparison of the adopted standard color excesses (see text) with reddening compilations by \citet[][top]{te87}, \citet[][middle]{ko08}, and \citet[][bottom]{an12}. The y-axis values are those of Table~\ref{tab1} supplemented by {\it KHG} reddenings in subsequent comparisons. Gray lines represent results expected for coincident reddening scales, and black lines represent fits to the data.}}
\label{fig2}
\end{figure}

The Cepheid space reddenings of Table~\ref{tab2} were first compared with the color excesses of \citet*{te87}, which were derived using reddening-independent {\it KHG} data from \citet{fm80} in conjunction with a small set of space reddenings by \citet{tu84}, where some of the latter differ slightly from the values presented here. There are 16 stars in common to the two data samples, the comparison being depicted in Fig.~\ref{fig2}~(top). The photometric {\it KHG} index of Brigham Young University is essentially independent of interstellar reddening \citep[see discussion by][]{te87}, and {\it KHG} reddenings depend upon Cepheid space reddenings only for calibration purposes. They therefore make a useful first test of the reddening compilation being established here.

A combination of least squares and non-parametric straight line fits to the data of Fig.~\ref{fig2}~(top) gave the formal results summarized in Table~\ref{tab3} and depicted in the figure. Despite small differences in the reddenings of the calibrating objects for the original {\it KHG} reddenings \citep{te87}, they appear to agree closely enough with the Table~\ref{tab2} data to be considered a reasonable match. Thereafter, the color excesses from \citet{te87} were used as published to augment the system of reddenings in Table~\ref{tab2}. The resulting standard system of 91 objects contains 52 stars from Table~\ref{tab2}, average values for 16 stars in common to the two systems, and 23 additional Cepheids from \citet{te87}. The system of standard reddenings denoted here as ``Std'' refers to the augmented system in the other comparisons made in Fig.~\ref{fig2}.

The reddening system of \citet{ko08} was expected to fall very close to the space and {\it KHG} reddening scales because of the manner in which the color excesses were derived relative to a stellar atmosphere model-based effective temperature scale. But differences were noted previously for those reddenings with respect to the empirical color-effective temperature scale of \citet{gr92} by \citet{te13a,te13b}, possibly arising from the surface gravity term included in the derivation of colors from effective temperatures by \citet{ko08}. A comparison of all overlapping reddenings in Fig.~\ref{fig2}~(middle) indicates a small zero-point offset and reddening-dependent term between the two scales, as noted by the fit in Table~\ref{tab3}.

The \citet{ko08} color excesses are readily tied to the system of space and {\it KHG} reddenings using the relationship of Table~\ref{tab3}, which was the procedure used here to augment the standard system of reddenings. There is moderate scatter in the reddening comparison, however, and it is not clear in which system it originates. About a third of the reddenings in the \citet{ko08} sample are tied to only 1 or 2 spectra, and that applies to the most deviant data in Fig.~\ref{fig2}~(middle). The remaining two thirds of the sample is tied to Cepheids with anywhere from 3 to 26 spectra, so lack of spectroscopic sampling cannot be the origin of the scatter. The adjusted reddenings from \citet{ko08} were therefore assimilated into the standard system with some care, particularly for deviant Cepheids in the comparison sample.

The reddening system of \citet{an12} was derived in analogous fashion to that of \citet{ko08}, so was expected to yield similar results, given its stellar atmosphere model-based effective temperature scale. All stars in their compilation were analyzed previously for reddening by \citet{ko08}. The comparison in Fig.~\ref{fig2}~(bottom) indicates both a zero-point offset and a larger reddening dependence for the \citet{an12} reddenings. Presumably both arise because the \citet{an12} reddening system was calibrated using an intrinsic color relation tied to both $T_{\rm eff}$ and $\log g$, as in \citet{ko08}, but normalized relative to the reddening scale of \citet{st11}. In order to agree with the present reddening scale, it is necessary to apply a zero-point offset of +0.05 and a scaling factor of 0.89 to the \citet{an12} color excesses, specifically by the relationship given in Table~\ref{tab3}. The scatter in the latter is slightly smaller than for the \citet{ko08} reddenings, most likely because they are restricted to Cepheids well-sampled spectroscopically.

The availability of two or more independent estimates of reddening for each Cepheid formed the rationale for creating a more extensive standard system of 198 Cepheids by averaging the adjusted color excesses of \citet{an12} with the space reddenings, {\it KHG} reddenings, and adjusted \citet{ko08} reddenings.

\begin{figure}[h]
\begin{center}
\includegraphics[width=0.45\textwidth]{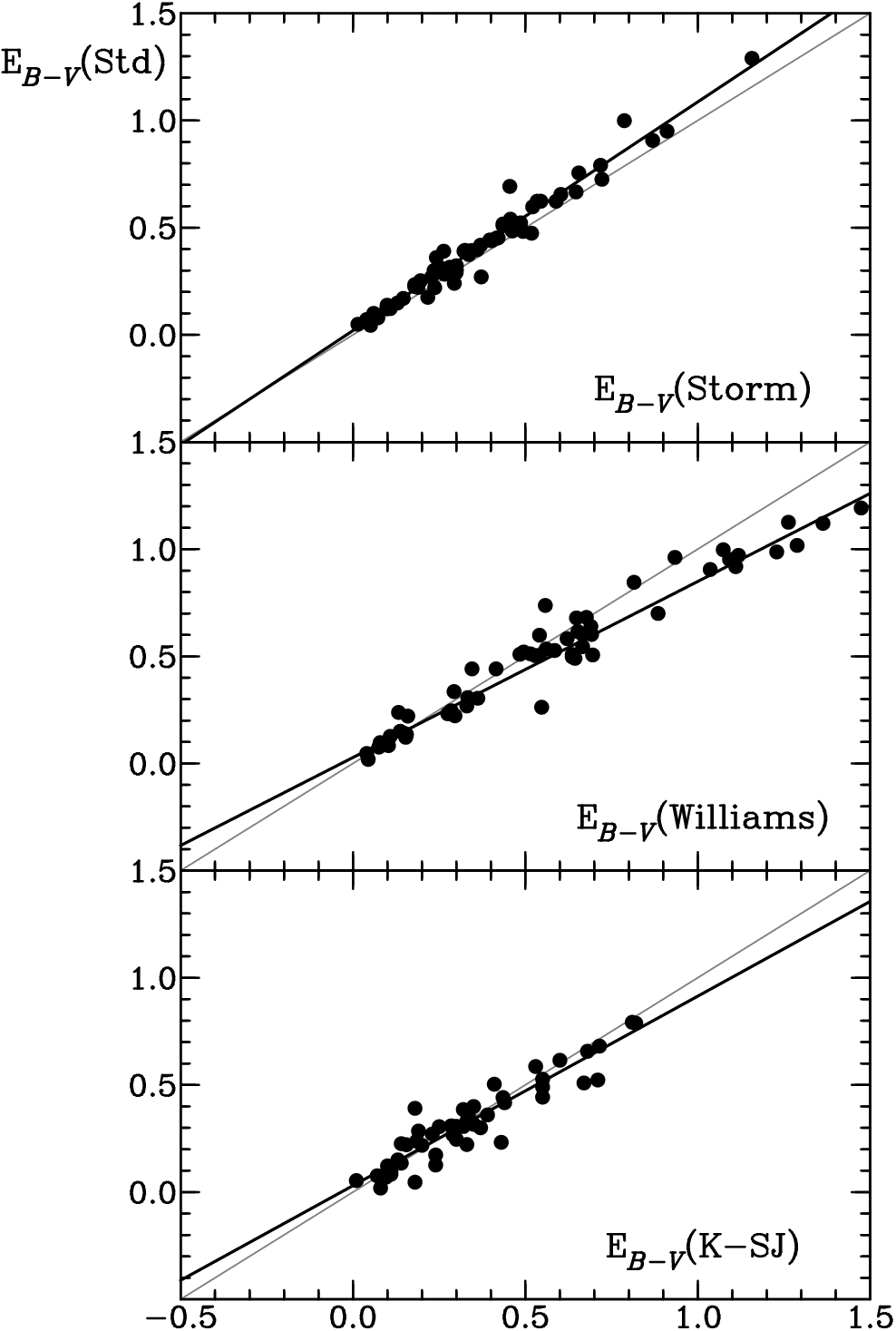}
\end{center}
\caption{\small{A similar comparison to that of Fig.~\ref{fig2} for x-axis color excesses from \citet[][top]{st11}, \citet[][middle]{wi66}, and an average of the reddenings from \citet{kr60} and \citet[][bottom]{sj89}, with y-axis values tied to the extended 200-star standard system. Black and gray lines are as in Fig.~\ref{fig2}.}}
\label{fig3}
\end{figure}

The reddening system of the \citet{st11} study was established by applying the infrared surface-brightness (IRSB) method to their Cepheid sample, with the \citet{ca89} extinction law used to adjust the observed star brightnesses in different wavelength bands. A comparison of the derived reddenings from that system with the standard system developed here is depicted in Fig.~\ref{fig3}~(top), and a regression fit is tabulated in Table~\ref{tab3}. The zero-point for the \citet{st11} system agrees closely with the present standard (Std) reddening system, but there is moderate scatter and a reddening-dependent trend opposite to that for the \citet{an12} system. The effect may be related to the manner of making extinction corrections, given that the \citet{ca89} extinction law has recently been shown to conflict with actual extinction parameters in many Galactic fields \citep{zg13,te14}. Large scatter for some Cepheids can be explained by the fact that they are southern hemisphere objects for which only \citet{ko08} reddenings are available for comparison, often Cepheids sampled by \citet{ko08} with only a single spectrum at one phase of the light curve.

An initial melding with the present system of the \citet{st11} reddenings, modified according to the relationship in Table~\ref{tab3}, created noticeable discrepancies for three stars. Two in particular, WZ~Car and S~Vul, have solid field reddenings in Table~\ref{tab2}, while the reddening for VZ~Pup differed by $\sim0^{\rm m}.2$ from the reddening derived by \citet{ko08}. Initially it was decided to include only the adjusted reddenings of \citet{st11} for the 57 least deviant objects in the extended standard system, and subsequent analyses were carried out accordingly. It was later noticed that the \citet{ko08} reddening for VZ~Pup created an anomalous intrinsic color for the variable, whereas that from the adjusted \citet{st11} reddening did not. VZ~Pup is a $23^{\rm d}$ Cepheid, and the long-period pulsators, in particular, evolve the fastest, display the most rapid rates of period change \citep{te06a}, and undergo the largest random period fluctuations \citep{te09c}. Given that the \citet{ko08} reddening for the star is tied to only one spectrum, there is a possibility that the derived light curve phase for that spectrum is erroneous, thereby affecting the reddening in systematic fashion. None of the other Cepheid reddening comparisons involve VZ~Pup in significant fashion, so the later decision to exclude the transformed \citet{ko08} reddening for the star in favor of the transformed \citet{st11} reddening does not impact the analyses.

\setcounter{table}{2}
\begin{table}[ht]
\caption[]{\small{Reddening Comparisons.}}
\label{tab3}
\centering
\small
\begin{tabular*}{0.48\textwidth}{@{\extracolsep{-2.7mm}}lccc}
\hline \hline \noalign{\medskip}
{\it X}-coordinate &{\it a} (zero-point) &{\it b} (slope) &n \\
\noalign{\smallskip} \hline \noalign{\medskip}
\noalign{\smallskip}
E$_{B-V}$({\it KHG}) &$-0.013 \pm0.009$ &$1.050 \pm0.049$ &16 \\
E$_{B-V}$(Kovtyukh) &$+0.037 \pm0.006$ &$0.974 \pm0.019$ &58 \\
E$_{B-V}$(Andrievsky) &$+0.053 \pm0.008$ &$0.890 \pm0.016$ &26 \\
E$_{B-V}$(Storm) &$+0.020 \pm0.006$ &$1.067 \pm0.019$ &64 \\
E$_{B-V}$(Williams) &$+0.028 \pm0.009$ &$0.822 \pm0.020$ &56 \\
E$_{B-V}$(K-SJ) &$+0.030 \pm0.013$ &$0.884 \pm0.038$ &48 \\
E$_{B-V}$(Mianes) &$+0.013 \pm0.026$ &$0.992 \pm0.063$ &19 \\
E$_{B-V}$(PBo) &$+0.006 \pm0.009$ &$0.995 \pm0.028$ &40 \\
E$_{B-V}$(PBe) &$+0.013 \pm0.009$ &$1.091 \pm0.034$ &43 \\
E$_{B-V}$(FM80) &$+0.019 \pm0.006$ &$0.861 \pm0.017$ &37 \\
E$_{B-V}$(Fe87) &$+0.045 \pm0.007$ &$0.887 \pm0.024$ &38 \\
E$_{B-V}$(Gray) &$+0.050 \pm0.004$ &$0.851 \pm0.015$ &40 \\
E$_{B-V}$(TY) &$+0.020 \pm0.009$ &$0.901 \pm0.019$ &105 \\
E$_{B-V}$(Harris) &$+0.082 \pm0.011$ &$0.878 \pm0.018$ &70 \\
E$_{B-V}$(Dean) &$+0.014 \pm0.013$ &$1.007 \pm0.038$ &20 \\
E$_{B-V}$(Pel) &$+0.077 \pm0.008$ &$0.876 \pm0.021$ &62 \\
E$_{B-V}$(Bersier) &$+0.045 \pm0.008$ &$0.849 \pm0.030$ &29 \\
E$_{B-V}$(Kiss) &$+0.066 \pm0.016$ &$0.738 \pm0.057$ &18 \\
E$_{B-V}$(Fernie67) &$-0.047 \pm0.012$ &$1.056 \pm0.027$ &111 \\
E$_{B-V}$(Fernie90) &$+0.053 \pm0.005$ &$0.922 \pm0.011$ &157 \\
E$_{B-V}$(Sasselov) &$+0.029 \pm0.013$ &$0.707 \pm0.096$ &5 \\
E$_{B-V}$(Krockenberger) &$+0.016 \pm0.008$ &$0.849 \pm0.038$ &11 \\
E$_{B-V}$(DWC) &$+0.017 \pm0.010$ &$0.988 \pm0.027$ &34 \\
E$_{B-V}$(LS) &$+0.051 \pm0.014$ &$0.900 \pm0.021$ &20 \\
E$_{B-V}$(LC) &$+0.032 \pm0.004$ &$1.000 \pm0.014$ &40 \\
\noalign{\smallskip} \hline
\end{tabular*}
\end{table}

The close agreement of the Table~\ref{tab2} reddenings with results obtained using the {\it KHG} index \citep{te87}, the reddening-independent parameter developed at Brigham Young University for measuring the strength of the stellar G band relative to the Ca II K line and $H\delta$, raises the question of whether similar results apply to reddenings derived in other systems employing a reddening-free parameter. The reddening system of \citet{wi66}, for example, was calibrated using cluster Cepheids and the break in the continuum across the {\it G} band in Cepheid spectra, in similar fashion to the method of establishing $T_{\rm eff}$ from {\it KHG} photometry. The Str\"{o}mgren system reddenings, E$_{b-y}$, of \citet{wi66} were converted to Johnson system reddenings, E$_{B-V}$, using the 0.73 factor adopted by \citet{te87}, and are compared with the present results in Fig.~\ref{fig3}~(middle), as well as in Table~\ref{tab3}. The scatter in the reddening comparison is large, making the \citet{wi66} color excesses of little value for averaging into our standard system. The scaling factor of 0.82 may relate to the manner in which the system was calibrated using the data inferred for cluster Cepheids in that era or how reddening was treated in the Str\"{o}mgren system.

$\Gamma$-index reddenings \citep{kr60,sj89}, which are also tied to the strength of the {\it G} band in stellar spectra, as measured photometrically, are presumably reddening-free as well. They were calibrated in the original studies using known reddenings for FG-type supergiants and cluster Cepheids. A comparison of our standard system reddenings with $\Gamma$-index color excesses is depicted in Fig.~\ref{fig3}~(bottom), where the $\Gamma$ reddenings from \citet{kr60} and \citet{sj89} were averaged in an attempt to reduce the rather sizeable scatter in each. A regression fit to the data is given in Table~\ref{tab3}. The $\Gamma$-index reddenings display large scatter no matter how the data are combined. They were therefore omitted from possible inclusion in an augmented standard system.

\begin{figure}[h]
\begin{center}
\includegraphics[width=0.45\textwidth]{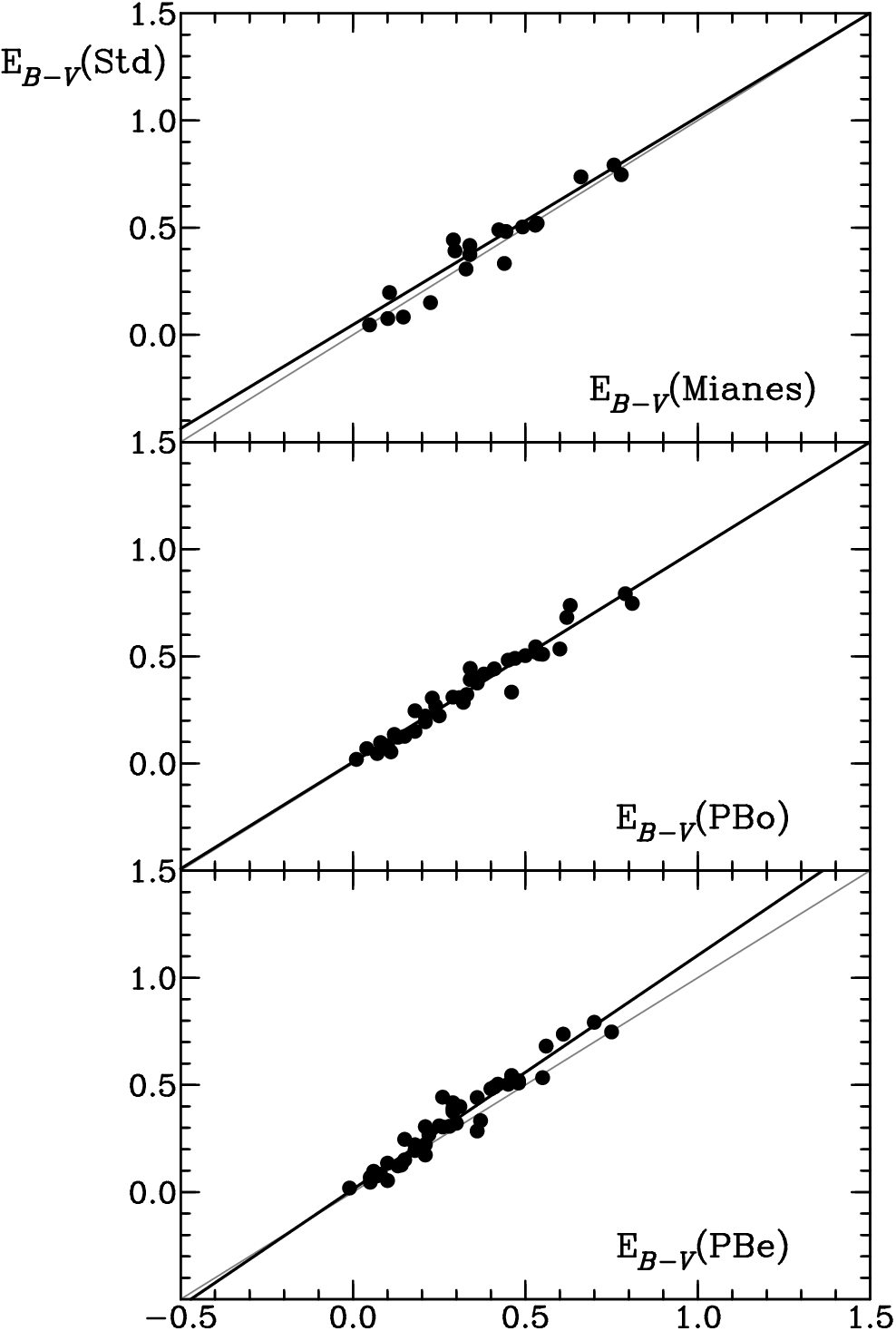}
\end{center}
\caption{\small{A similar comparison to that of Figs.~\ref{fig2}~\&~\ref{fig3} for color excess compilations by \citet[][top]{mi63}, \citet[][middle]{pb71}, and \citet[][bottom]{pb75}. Lines are as in Fig.~\ref{fig2}.}}
\label{fig4}
\end{figure}

The comparisons so far have been to systems that should be independent of any period-color formulation. That is not always the case for other published reddening systems. Three comparisons with less well-defined systems are presented in Fig.~\ref{fig4}, and summarized in Table~\ref{tab3}. All three systems are tied to Lick six-color ({\it UVBGRI}) photometry \citep{kr58,sk64} of Cepheids by \citet{mi63} and the sources cited by \citet{pa71}. Fig.~\ref{fig4}~(top) compares the reddenings of \citet{mi63} with the present standard system, the former being converted from {\it G--I} reddenings to {\it B--V} reddenings using the 1.89 conversion factor of \citet{sc69}. The \citet{mi63} reddenings appear to agree overall with the color excesses of Table~\ref{tab2}, but the scatter for individual Cepheids is too large to make them suitable for incorporation into the standard system.

Fig.~\ref{fig4}~(middle) is a comparison with the reddenings of \citet{pb71}, derived by means of model atmosphere and stellar evolutionary model fits to Lick six-color photometry of Cepheids. Fig.~\ref{fig4}~(bottom) presents similar results for the subsequent study by \citet{pb75} using a similar methodology with a {\it Q} parameter. So-called ``reddening-free'' {\it Q} parameters depend directly on the reddening law adopted, which, as noted in \S2, is variable around the Galaxy. Reddenings tied to {\it Q} parameters are therefore inherently unreliable, and that applies here to the \citet{pb75} system linked to Lick six-color photometry.

The reddenings of \citet{pb75} and \citet{pb71} have long been considered to be of high quality, although there is a negative offset of the former relative to the latter, presumably arising from the different methods of analysis. The generally small scatter in Fig.~\ref{fig4}~(middle) and the small uncertainties in the regression fit (Table~\ref{tab3}) confirm the relatively high quality of the \citet{pb71} reddenings, despite uncertainties in how interstellar reddening was treated. The Cepheids in the \citet{pb71} sample were therefore included in the standard system without adjustment, although the residual scatter is of concern. Similar statements do not apply to the \citet{pb75} reddenings, which display slightly larger scatter and a noticeable reddening-dependent trend. It was decided to omit them from inclusion in the standard system.

\begin{figure}[h]
\begin{center}
\includegraphics[width=0.45\textwidth]{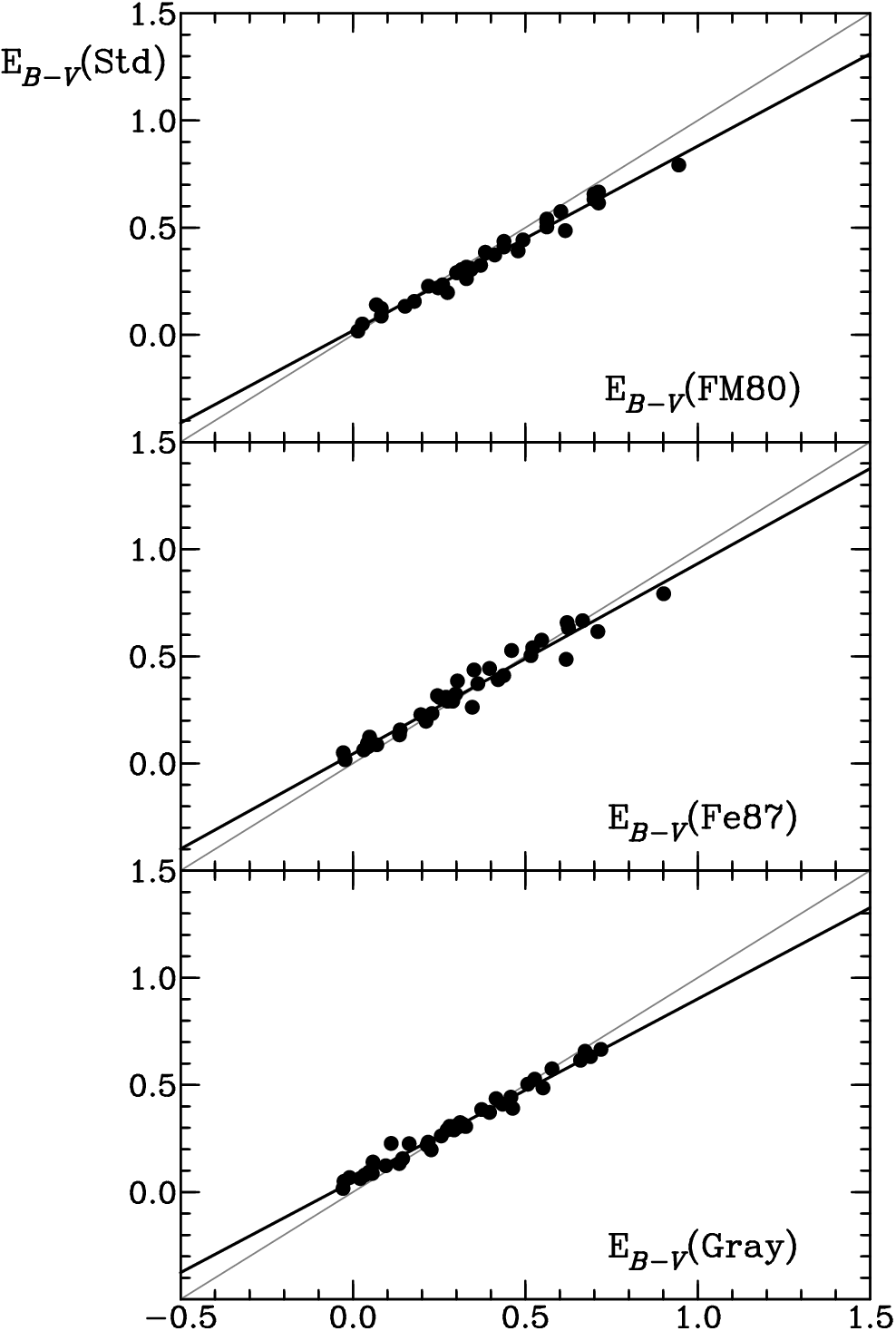}
\end{center}
\caption{\small{A similar comparison to that of Figs.~\ref{fig2}~\&~\ref{fig3} for color excess compilations by \citet[][top]{fm80}, \citet[][middle]{fe87}, and \citet[][bottom]{gr91}. Lines are as in Fig.~\ref{fig2}.}}
\label{fig5}
\end{figure}

Fig.~\ref{fig5} presents comparisons with the present standard system for three reddening sources on the Str\"{o}mgren system, all derived from the observations of \citet{fm80}: the original color excesses of \citet[][top]{fm80}, those of \citet[][middle]{fe87}, and those of \citet[][bottom]{gr91}. As was the case for the reddenings of \citet{wi66}, the Str\"{o}mgren system reddenings E$_{b-y}$ were converted to Johnson system reddenings E$_{B-V}$ using the factor of 0.73 adopted by \citet{te87}. All three sources compared in Fig.~\ref{fig5} and summarized in Table~\ref{tab3} appear to display reddening-dependent offsets that likely originate from the adoption of reddening-free indices for Str\"{o}mgren photometry that are not strictly ``reddening-free'' according to the arguments presented above. Despite that, the scatter about the best-fitting relations is generally small, although that for the \citet{gr91} reddenings is sufficiently tight to allow them to be amalgamated into the present standard system after adjustment for a zero-point offset of +0.05 and a scaling factor of 0.85. The actual adjustments were made using the Table~\ref{tab3} relationship.

\begin{figure}[h]
\begin{center}
\includegraphics[width=0.45\textwidth]{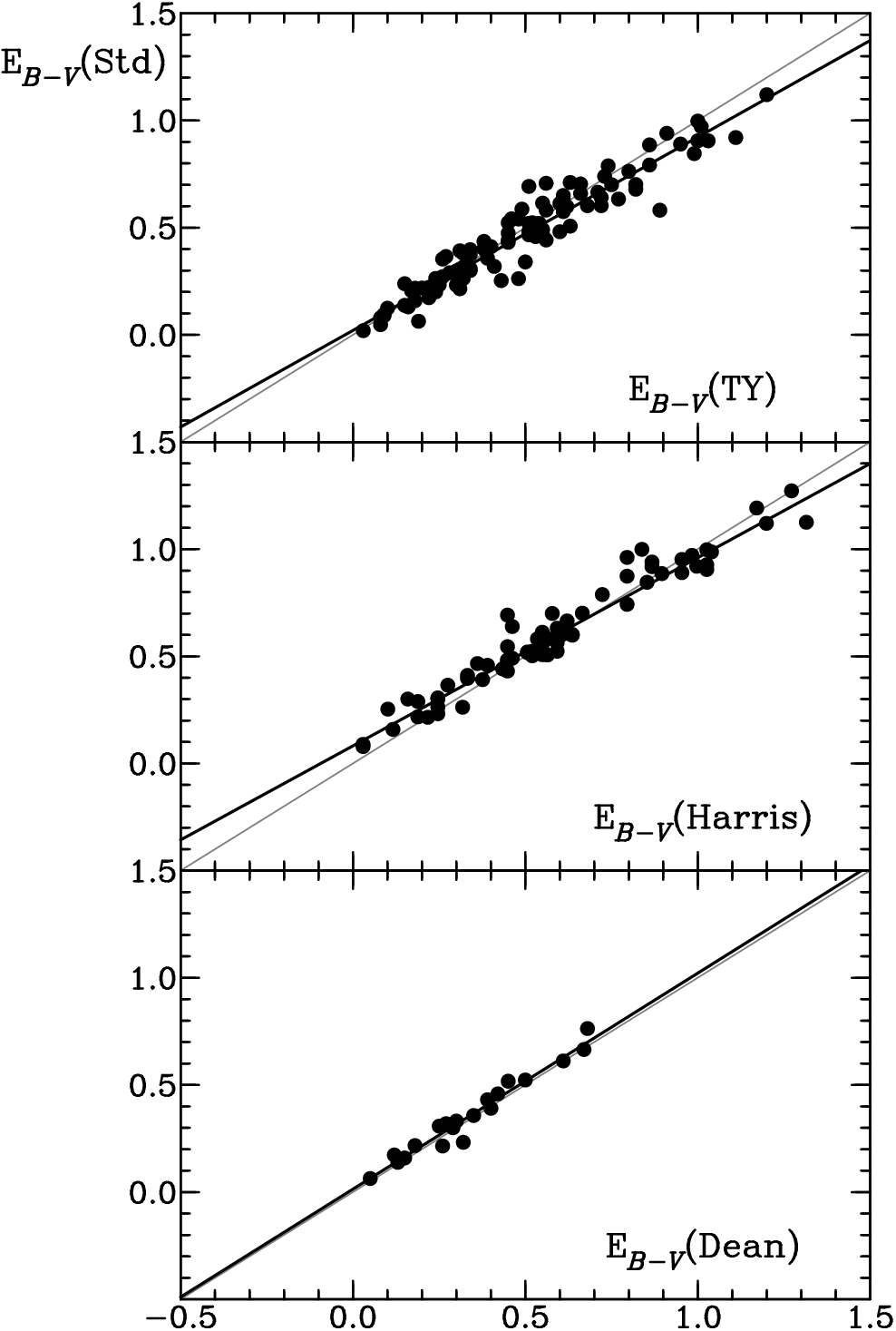}
\end{center}
\caption{\small{A similar comparison to that of Figs.~\ref{fig2}~\&~\ref{fig3} for color excess compilations by \citet[][top]{ty70}, \citet[][middle]{ha81a,ha81b}, and \citet[][bottom]{de81}. Lines are as in Fig.~\ref{fig2}.}}
\label{fig6}
\end{figure}

Fig.~\ref{fig6} and Table~\ref{tab3} present comparisons with the present system for three reddening sources tied to spectroscopic observations of Cepheids, either mixed spectroscopy/photometry as in the case of the color excesses by \citet[][Fig.~\ref{fig6}~(top)]{ty70} or using photometric systems designed to make use of gravity and temperature-sensitive features in the spectra of yellow and red supergiants: the Washington system used by \citet[][Fig.~\ref{fig6}~(middle)]{ha81a,ha81b}, with conversion from E$_{T_1-T_2}$ to E$_{B-V}$ following \citet{ha81a}, and the DDO system used by \citet[][Fig.~\ref{fig6}~(bottom)]{de81}. The Cepheid reddening system of \citet{ha81a,ha81b} was calibrated using a period-color relation tied to cluster members, so the large scatter in the resulting reddenings is accounted for by the fact that Cepheids of identical period lying in the instability strip can have appreciably different effective temperatures, hence colors. The large scatter for the reddenings of \citet{ty70} may be of similar origin, given that they are tied to the spectroscopic inferences of Kraft, some of which are linked to $\Gamma$-index reddenings \citep{kr60}, previously indicated to exhibit large scatter (Fig.~\ref{fig3}). The DDO system reddenings of \citet{de81} are a remarkably close fit to the present system reddenings, however, and were therefore amalgamated into the latter as cited.

\begin{figure}[h]
\begin{center}
\includegraphics[width=0.45\textwidth]{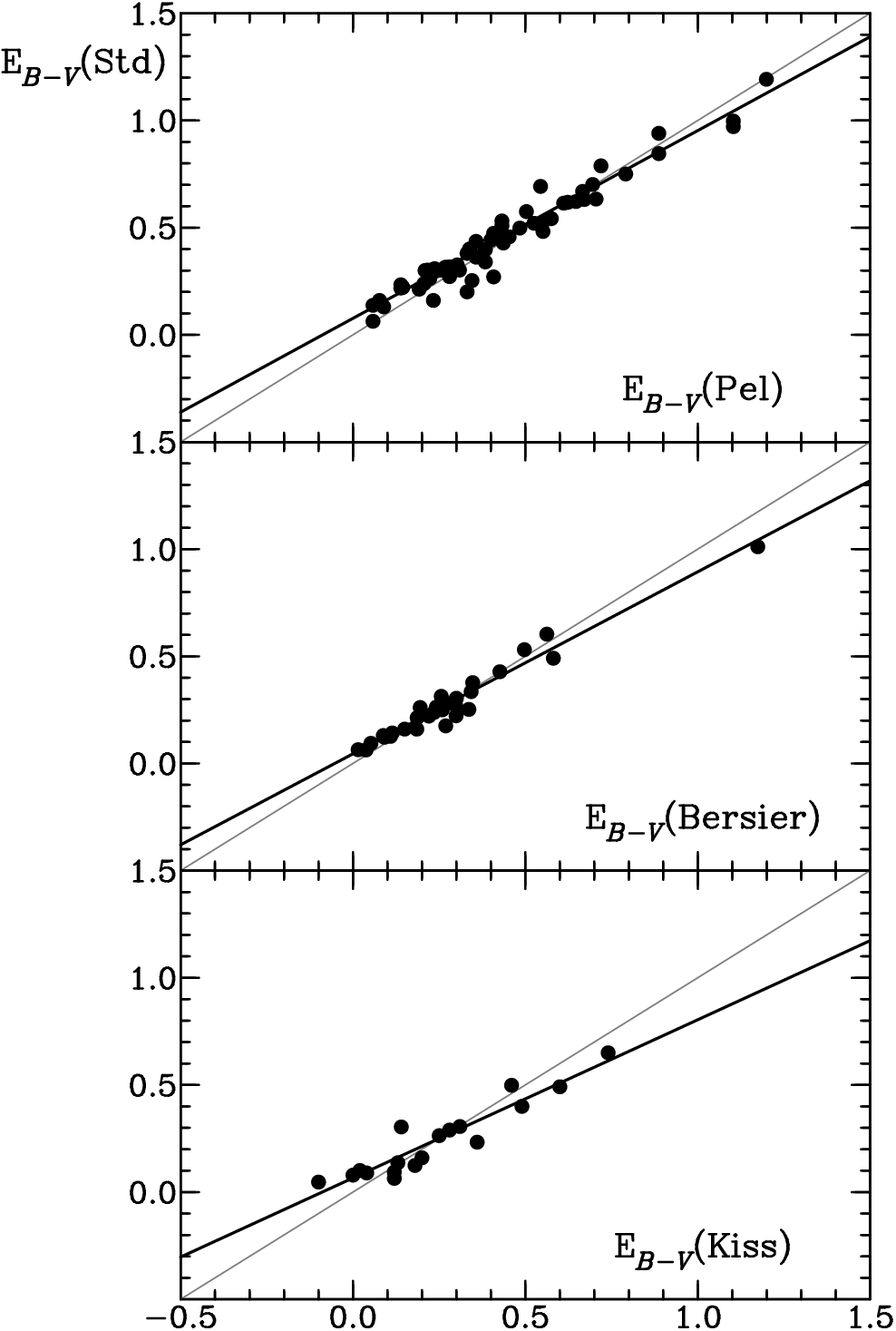}
\end{center}
\caption{\small{A similar comparison to that of Figs.~\ref{fig2}~\&~\ref{fig3} for color excess compilations by \citet[][top]{pe78}, \citet[][middle]{br96}, and \citet[][bottom]{ks98}. Lines are as in Fig.~\ref{fig2}.}}
\label{fig7}
\end{figure}

Three more reddening sources are compared with the present standard system in Fig.~\ref{fig7}, with regression fits summarized in Table~\ref{tab3}. Fig.~\ref{fig7}~(top) shows a comparison with the reddenings of \citet{pe78} on the Walraven photometric system, converted to Johnson system reddenings using the relationship adopted by \citet{pe85}. Fig.~\ref{fig7}~(middle) shows a comparison with the Geneva system reddenings of \citet{br96}, while Fig.~\ref{fig7}~(bottom) is a comparison with the mixed-source reddenings of \citet{ks98}. The Walraven system reddenings \citep{pe78} do not match the present standard system, displaying large scatter as well as zero-point and scaling offsets. The reasons are unclear.

The Geneva system reddenings \citep{br96} are a much better match to the present system, but with sufficient scatter to make the choice of amalgamation into the standard set of this paper a difficult decision. Six Cepheids in particular do not fit the general trend, and again the reason is uncertain. Meanwhile, the final comparison shown in Fig.~\ref{fig7} speaks to the unusual method of calibration used by \citet{ks98} in their study. The large scatter, as well as the zero-point offset and large scaling term, indicate a very poor match to the present standard system. The color excesses inferred in the \citet{ks98} study should not be used for studies of the intrinsic properties of Cepheids.

\begin{figure}[h]
\begin{center}
\includegraphics[width=0.45\textwidth]{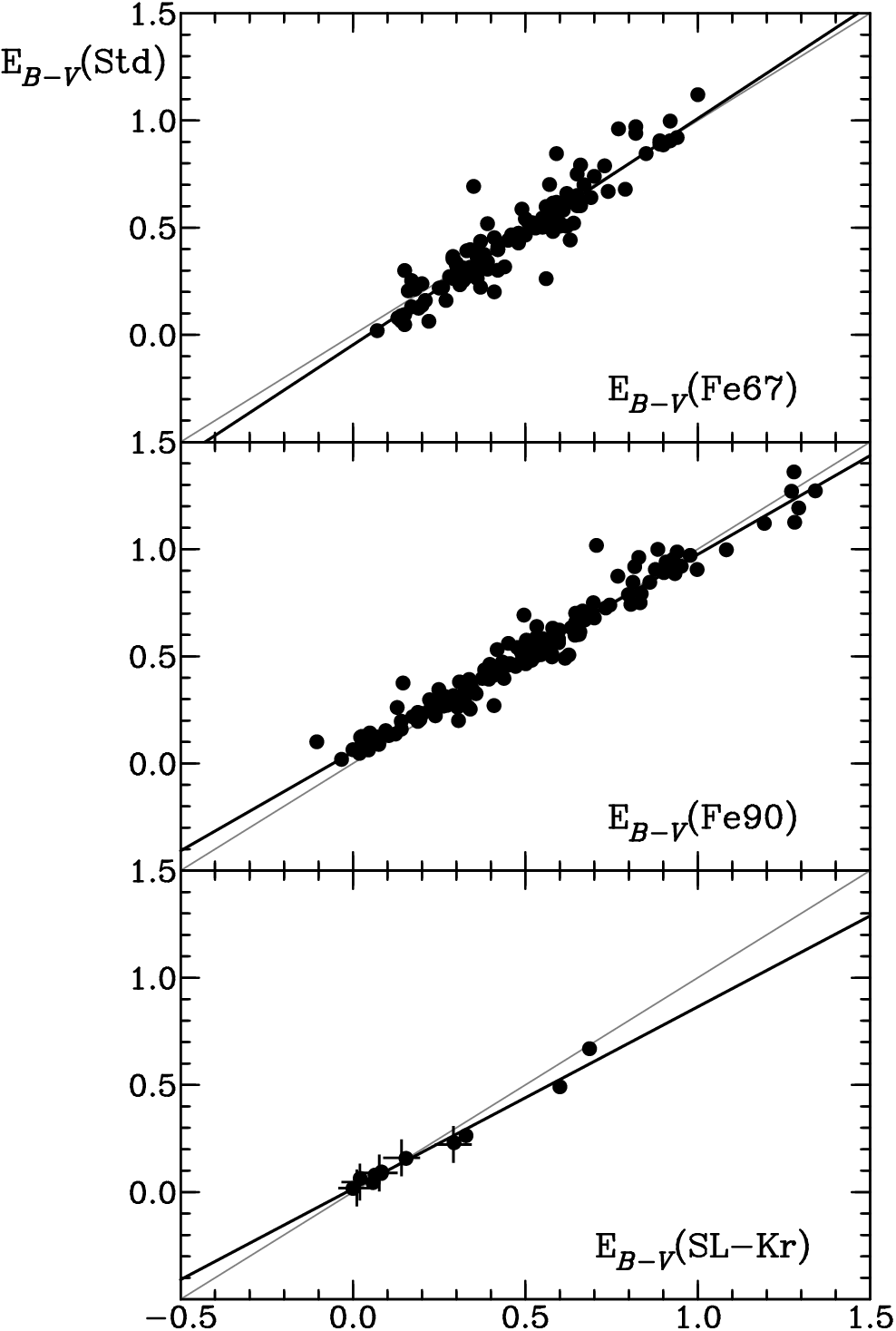}
\end{center}
\caption{\small{A similar comparison to that of Figs.~\ref{fig2}~\&~\ref{fig3} for color excess compilations by \citet[][top]{fe67}, \citet[][photometric scale, middle]{fe90a}, and \citet[][bottom, plus signs]{sl90} and \citet[][bottom]{kr98}. Lines are as in Fig.~\ref{fig2}.}}
\label{fig8}
\end{figure}

Fig.~\ref{fig8} and Table~\ref{tab3} show comparisons with two large compilations of Cepheid reddenings: an older but useful study by \citet{fe67} presented in Fig.~\ref{fig8}~(top), and the oft-cited homogeneous collection of \citet[][photometric scale]{fe90a} presented in Fig.~\ref{fig8}~(middle). The bottom portion of the plot is a comparison with the reddenings of 5 Cepheids derived by \citet{sl90} and 11 Cepheids (including the previous 5) derived by \citet{kr98} using line ratios for infrared spectroscopic lines such as C~{\small I} and Si~{\small I} to infer effective temperatures and, hence, intrinsic colors. Both Fernie compilations went to some length to place reddenings derived in various studies on a homogeneous system, and both scales appear to be a reasonably good match with the system of space reddenings for Cepheids presented here. Reddening-dependent trends in the two compilations appear to be negligible. But the scatter is large in both cases, more so in the case of the older study by \citet{fe67}, so they cannot be adopted as published for amalgamation with the present standard system. Similar results apply to the cluster reddening scale of \citet{fe90a}.

The comparison with the results of \citet{sl90} and \citet{kr98} in Fig.~\ref{fig8}~(bottom) is useful for indicating that the premise of using infrared spectroscopic line ratios for Cepheids to infer redddenings is quite sound, as also demonstrated by \citet{ko08} and \citet{an12}. The reddenings for four of the five Cepheids studied by \citet{sl90}, plotted using plus signs, agree closely with the present results. The fifth Cepheid, T~Mon, is somewhat discrepant with respect to most photometric reddening estimates for the star, but does follow the trend for other stars with reddenings derived in similar fashion \citep{kr98}. The discussion by \citet{sl90} notes the difference with respect to the \citet{fe90a} reddening scale for T~Mon. The field of this Cepheid is not well-studied with regard to the reddening of nearby companions, so it may be possible to resolve the minor discrepancy for the star with additional observations of nearby companions to T~Mon. Given the generally small scatter for the reddenings, the color excesses of \citet{sl90} and \citet{kr98} were included in the standard system of this paper using the adjustments cited in Table~\ref{tab3} for the \citet{kr98} results.

\begin{figure}[h]
\begin{center}
\includegraphics[width=0.45\textwidth]{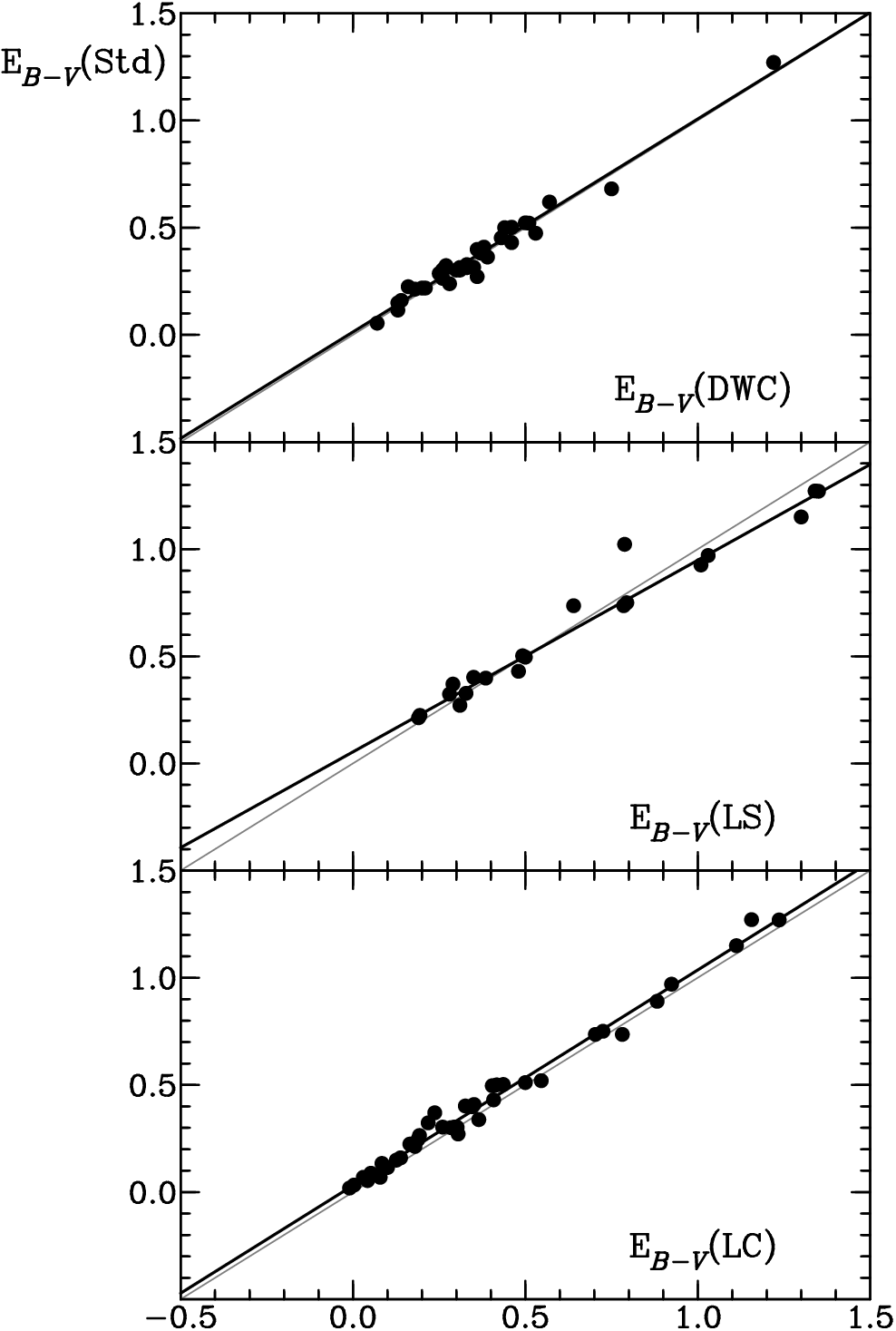}
\end{center}
\caption{\small{A similar comparison to that of Figs.~\ref{fig2}~\&~\ref{fig3} for color excess compilations by \citet[][top]{de78}, \citet[][middle]{ls93}, and \citet[][bottom]{lc07}. Lines are as in Fig.~\ref{fig2}.}}
\label{fig9}
\end{figure}

Finally, Fig.~\ref{fig9} and Table~\ref{tab3} present comparisons with reddening systems tied to near infrared photometry of Cepheids. The study by \citet*{de78} using Cape system {\it BVI} photometry generated Cepheid reddenings that are well-matched to the standard system presented here, as indicated in Fig.~\ref{fig9}~(top). They were therefore amalgamated into the present standard system with only minor concerns about the small discrepancies for some stars.

The study by \citet{ls93} focused mainly on reddenings inferred from {\it JHK} observations of Cepheids, and was tied to a sequence of published reddenings for Galactic calibrators. The comparison is depicted in Fig.~\ref{fig9}~(middle). The most deviant data point corresponds to the Cepheid S~Vul, which did not have a solid field reddening in the era when the study was completed. Otherwise, the data display both a zero-point offset and a reddening dependence that conflict with the present system of reddenings. Residual scatter in some cases may be tied to other calibrators for which field reddenings were not well-established at the time of the \citet{ls93} study.

The study by \citet{lc07} was intended to solidify the Cape system reddenings with reference to Cepheid field reddenings, and the generally small scatter in the comparison displayed in Fig.~\ref{fig9}~(bottom) implies a very good match to the standard system established here, except for a noticeable zero-point offset. The relationship of Table~\ref{tab3} found for the \citet{lc07} reddenings was therefore used to adjust them to the present scale, in which they were subsequently included.

Reddenings have been derived by \citet{sc92} for several Cepheids using {\it JHK} photometry, but there is no overlap of their sample, all southern hemisphere objects, with the present selection of calibrating Cepheids. It is therefore not possible to test the manner of deriving Cepheid reddenings from period-color relations using {\it JHK} colors, although presumably the smaller scatter in intrinsic colors for Cepheids observed at far infrared wavelengths should overcome most of the objections that would apply if the same methodology had adopted optical band colors such as {\it B--V}. 

\section{{\rm \footnotesize SUMMARY}}
This study presents a collection of space reddenings for 68 Cepheids, and uses the sample to test a variety of published reddening compilations for Cepheids. Initial testing was made relative to the {\it KHG} reddenings of \citet{te87}, since the Brigham Young University {\it KHG} index is essentially reddening independent. Similar comparisons with the spectroscopic reddenings of \citet{ko08} and \citet{an12} revealed small zero-point and reddening-dependent offsets relative to the system of space and {\it KHG} reddenings. Corrections for those, and use of the \citet{te87} reddenings as published, permitted the creation of an expanded standard system comprising the best of the available reddenings adjusted to the space reddening and {\it KHG} reddening combination. The expanded standard system (Std) was used to make comparisons with more extensive systems of published reddenings, with the results presented in \S3 and Figs.~\ref{fig2}--\ref{fig9}.

\begin{figure}[!t]
\begin{center}
\includegraphics[width=0.45\textwidth]{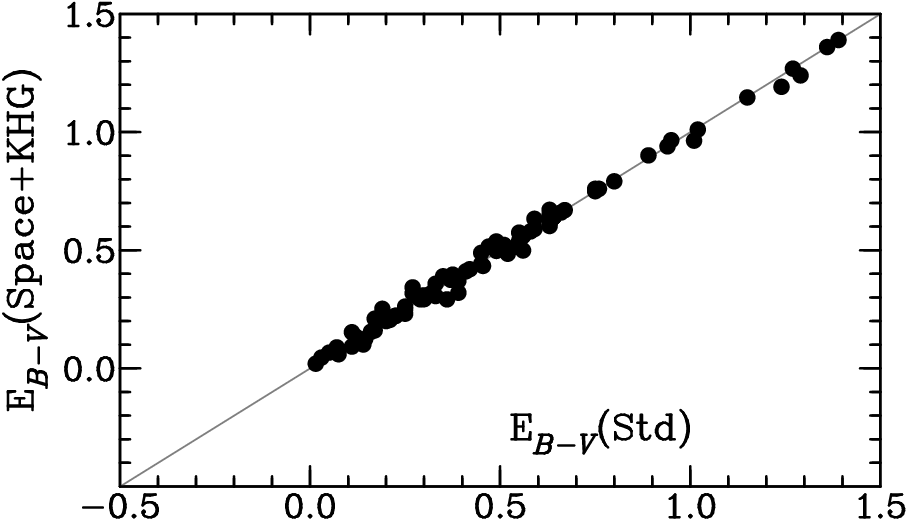}
\end{center}
\caption{\small{A comparison of the combined space and {\it KHG} reddenings ({\it y}-axis) with the system of standard reddenings developed in this study ({\it x}-axis). The gray line represents coincident reddening scales.}} 
\label{fig10}
\end{figure}

The comparative best of the other reddening systems sampled, once adjusted for zero-point and scaling effects, are those of \citet{ko08}, \citet{an12}, \citet{st11}, \citet{pb71}, \citet{gr91}, \citet{de81}, \citet{sl90}, \citet{kr98}, \citet{de78}, and \citet{lc07}. They are combined with the space reddenings of Table~\ref{tab2} and the {\it KHG} reddenings of \citet{te87} into the standard system of reddenings summarized in Table~\ref{tab4}, with the result for VZ~Pup modified as discussed earlier. A comparison of those reddenings with the original system of space and {\it KHG} reddenings is presented in Fig.~\ref{fig10}. There is clearly some residual scatter in the comparison, indicating the existence of uncertainties for some stars. Note, in particular, the derived standard reddenings of E$_{B-V} = 0.32$ and 0.37 for QZ~Nor and V340~Nor, respectively, relative to the common space reddenings of 0.39 adopted in \S2. It implies the necessity for additional work on problem objects to resolve the small residual differences. The average standard deviation for the combined reddenings of Cepheids with several independent estimates is $\pm0.028$, although for individual objects the values range from $\pm0.002$ to $\pm0.093$. Readers should therefore use caution when using the reddenings of Table~\ref{tab4} to deduce intrinsic properties for specific Cepheids.

\setcounter{table}{3}
\begin{table*}[htp]
\caption[]{\small{Reddening Summary.}}
\label{tab4}
\centering
\tiny
\begin{tabular*}{1.00\textwidth}{@{\extracolsep{-2.1mm}}lcccclcccclcccclccc}
\hline
Cepheid &$\log P$ &E$_{B-V}$ &({\it B--V})$_0$ & &Cepheid &$\log P$ &E$_{B-V}$ &({\it B--V})$_0$ & &Cepheid &$\log P$ &E$_{B-V}$ &({\it B--V})$_0$ & &Cepheid &$\log P$ &E$_{B-V}$ &({\it B--V})$_0$ \\
\hline
T~Ant &0.7707 &0.300 &0.407 & &CEa~Cas &0.7111 &0.508 &0.613 & &BG~Lac &0.7269 &0.322 &0.647 & &HW Pup &1.1289 &0.724 &0.518 \\
U~Aql &0.8466 &0.397 &0.629 & &CEb~Cas &0.6512 &0.497 &0.534 & &V473~Lyr &0.3225 &0.126 &0.457 & &LR~Pup &0.5226 &0.420 &0.510 \\
SZ~Aql &1.2340 &0.614 &0.797 & &CF~Cas &0.6880 &0.537 &0.658 & &T~Mon &1.4318 &0.223 &0.950 & &MY~Pup &0.7555 &0.142 &0.506 \\
TT~Aql &1.1385 &0.485 &0.808 & &CG~Cas &0.6401 &0.640 &0.580 & &SV~Mon &1.1828 &0.267 &0.776 & &V620~Pup &0.4126 &0.640 &0.520 \\
FF~Aql &0.6504 &0.231 &0.524 & &CH~Cas &1.1786 &0.920 &0.729 & &TW~Mon &0.8511 &0.701 &0.636 & &S~Sge &0.9234 &0.135 &0.667 \\
FM~Aql &0.7863 &0.633 &0.638 & &CY~Cas &1.1577 &0.905 &0.761 & &TX~Mon &0.9396 &0.482 &0.631 & &GY~Sge &1.7081 &1.147 &1.140 \\
FN~Aql &0.9769 &0.521 &0.695 & &DD~Cas &0.9918 &0.465 &0.766 & &TZ~Mon &0.8709 &0.473 &0.655 & &U~Sgr &0.8290 &0.434 &0.666 \\
V496~Aql &0.8330 &0.442 &0.715 & &DL~Cas &0.9031 &0.523 &0.630 & &UY~Mon &0.3799 &0.153 &0.389 & &W~Sgr &0.8805 &0.133 &0.612 \\
V600~Aql &0.8597 &0.845 &0.684 & &FM~Cas &0.7641 &0.353 &0.635 & &WW~Mon &0.6686 &0.707 &0.426 & &X~Sgr &0.8459 &0.252 &0.502 \\
V733~Aql &0.7909 &0.140 &0.786 & &FO~Cas &0.8324 &0.760 &0.605 & &XX~Mon &0.7369 &0.631 &0.531 & &Y~Sgr &0.7614 &0.222 &0.635 \\
V1162~Aql &0.7305 &0.196 &0.685 & &IO~Cas &0.7485 &0.590 &0.572 & &AA~Mon &0.5953 &0.742 &0.589 & &VY~Sgr &1.1322 &1.192 &0.798 \\
V1344~Aql &0.8738 &0.567 &0.815 & &V379~Cas &0.6341 &0.621 &0.525 & &AC~Mon &0.9039 &0.575 &0.605 & &WZ~Sgr &1.3395 &0.499 &0.884 \\
$\eta$~Aql &0.8559 &0.155 &0.634 & &V636~Cas &0.9231 &0.603 &0.779 & &CU~Mon &0.6728 &0.874 &0.525 & &XX~Sgr &0.8078 &0.586 &0.562 \\
V340~Ara &1.3183 &0.563 &1.008 & &V~Cen &0.7399 &0.305 &0.576 & &CV~Mon &0.7307 &0.760 &0.545 & &YZ~Sgr &0.9802 &0.319 &0.716 \\
RT~Aur &0.5715 &0.081 &0.510 & &V810~Cen &2.1844 &0.250 &0.550 & &EE~Mon &0.6821 &0.533 &0.480 & &AP~Sgr &0.7040 &0.217 &0.587 \\
RX~Aur &1.0654 &0.301 &0.657 & &CP~Cep &1.2519 &0.679 &0.955 & &EK~Mon &0.5975 &0.542 &0.647 & &AV~Sgr &1.1879 &1.125 &0.950 \\
SY~Aur &1.0062 &0.466 &0.539 & &CR~Cep &0.7947 &0.739 &0.655 & &FG~Mon &0.6529 &0.646 &0.541 & &AY~Sgr &0.8175 &0.940 &0.564 \\
YZ~Aur &1.2599 &0.602 &0.756 & &IR~Cep &0.3251 &0.395 &0.521 & &FI~Mon &0.5169 &0.628 &0.488 & &BB~Sgr &0.8220 &0.308 &0.662 \\
AN~Aur &1.0124 &0.581 &0.639 & &V351~Cep &0.4481 &0.462 &0.523 & &V465~Mon &0.4335 &0.175 &0.565 & &V350~Sgr &0.7122 &0.318 &0.588 \\
BK~Aur &0.9032 &0.423 &0.636 & &$\delta$~Cep &0.7297 &0.089 &0.572 & &V504~Mon &0.4431 &0.610 &0.394 & &V1882~Sgr &0.4332 &0.640 &0.530 \\
CY~Aur &1.1414 &0.918 &0.663 & &BG~Cru &0.5241 &0.122 &0.496 & &V508~Mon &0.6163 &0.252 &0.627 & &RV~Sco &0.7826 &0.376 &0.587 \\
ER~Aur &1.1958 &0.639 &0.516 & &X~Cyg &1.2145 &0.262 &0.875 & &V526~Mon &0.4273 &0.249 &0.338 & &RY~Sco &1.3078 &0.788 &0.716 \\
EW~Aur &0.4248 &0.580 &0.490 & &SU~Cyg &0.5850 &0.125 &0.446 & &S~Nor &0.9892 &0.210 &0.738 & &KQ~Sco &1.4577 &0.927 &1.079 \\
V335~Aur &0.5331 &0.720 &0.412 & &SZ~Cyg &1.1793 &0.598 &0.881 & &TW~Nor &1.0329 &1.240 &0.761 & &V500~Sco &0.9693 &0.622 &0.652 \\
RX~Cam &0.8983 &0.556 &0.640 & &TX~Cyg &1.1676 &1.120 &0.666 & &QZ~Nor &0.7333 &0.320 &0.585 & &Z Sct &1.1106 &0.582 &0.747 \\
TV~Cam &0.7239 &0.506 &0.629 & &VX~Cyg &1.3039 &0.845 &0.879 & &V340~Nor &1.0526 &0.371 &0.790 & &RU~Sct &1.2945 &0.966 &0.690 \\
AB~Cam &0.7625 &0.700 &0.494 & &VY~Cyg &0.8953 &0.659 &0.558 & &Y~Oph &1.2336 &0.671 &0.681 & &SS~Sct &0.5648 &0.381 &0.573 \\
AD~Cam &1.0516 &0.952 &0.618 & &VZ~Cyg &0.6870 &0.292 &0.585 & &BF~Oph &0.6094 &0.289 &0.591 & &UZ~Sct &1.1686 &0.997 &0.897 \\
OX~Cam &0.7050 &0.750 &0.430 & &BZ Cyg &1.0061 &0.886 &0.713 & &RS~Ori &0.8789 &0.436 &0.511 & &EV~Sct &0.4901 &0.710 &0.447 \\
RY~CMa &0.6701 &0.269 &0.577 & &CD~Cyg &1.2323 &0.502 &0.766 & &GQ~Ori &0.9353 &0.302 &0.673 & &V367~Sct &0.7989 &1.269 &0.589 \\
TW~CMa &0.8448 &0.340 &0.631 & &DT~Cyg &0.3978 &0.060 &0.480 & &SV~Per &1.0465 &0.397 &0.619 & &ST~Tau &0.6058 &0.335 &0.514 \\
VZ~CMa &0.6484 &0.560 &0.420 & &MW~Cyg &0.7749 &0.711 &0.631 & &UX~Per &0.6595 &0.545 &0.499 & &SZ~Tau &0.6515 &0.308 &0.539 \\
XZ~CMa &0.4079 &0.310 &0.490 & &V386~Cyg &0.7208 &0.905 &0.588 & &UY~Per &0.7296 &0.902 &0.612 & &AE~Tau &0.5907 &0.598 &0.519 \\
YZ~CMa &0.4992 &0.560 &0.560 & &V402~Cyg &0.6400 &0.393 &0.612 & &VX~Per &1.0370 &0.481 &0.671 & &EF~Tau &0.5376 &0.388 &0.531 \\
CN~CMa &0.4059 &0.630 &0.510 & &V532~Cyg &0.5164 &0.564 &0.470 & &AS~Per &0.6966 &0.704 &0.596 & &EU~Tau &0.3227 &0.261 &0.421 \\
U~Car &1.5885 &0.295 &0.887 & &V924~Cyg &0.7460 &0.264 &0.578 & &AW~Per &0.8105 &0.540 &0.514 & &$\alpha$~UMi &0.5988 &0.020 &0.328 \\
SX~Car &0.6866 &0.200 &0.727 & &V1154~Cyg &0.6925 &0.265 &0.648 & &BM~Per &1.3608 &0.987 &0.803 & &T~Vel &0.6665 &0.331 &0.597 \\
VY~Car &1.2773 &0.292 &0.869 & &V1334~Cyg &0.5228 &0.101 &0.406 & &HQ~Per &0.9364 &0.566 &0.664 & &RY~Vel &1.4493 &0.609 &0.770 \\
WZ~Car &1.3620 &0.270 &0.886 & &V1726~Cyg &0.6270 &0.358 &0.554 & &HZ~Per &1.0523 &1.360 &0.767 & &RZ~Vel &1.3096 &0.331 &0.790 \\
GT~Car &1.1193 &0.660 &0.796 & &TX~Del &0.7900 &0.253 &0.483 & &MM~Per &0.6147 &0.505 &0.590 & &SW~Vel &1.3700 &0.390 &0.761 \\
$\ell$~Car &1.5507 &0.161 &1.101 & &$\beta$~Dor &0.9931 &0.067 &0.733 & &OT~Per &1.4165 &1.390 &0.867 & &SX~Vel &0.9800 &0.308 &0.580 \\
RW~Cas &1.1701 &0.441 &0.760 & &W~Gem &0.8984 &0.306 &0.607 & &V440~Per &0.8791 &0.313 &0.561 & &CS~Vel &0.7712 &0.754 &0.600 \\
RY~Cas &1.0841 &0.602 &0.769 & &AD~Gem &0.5784 &0.197 &0.497 & &X~Pup &1.4143 &0.451 &0.776 & &S~Vul &1.8355 &0.999 &0.891 \\
SU~Cas &0.4401 &0.293 &0.408 & &DX~Gem &0.4966 &0.531 &0.408 & &RS~Pup &1.6169 &0.519 &0.918 & &T~Vul &0.6469 &0.092 &0.544 \\
SW~Cas &0.7357 &0.541 &0.539 & &$\zeta$~Gem &1.0065 &0.046 &0.751 & &VZ~Pup &1.3649 &0.505 &0.656 & &U~Vul &0.9026 &0.650 &0.626 \\
SY~Cas &0.8097 &0.474 &0.501 & &V~Lac &0.6975 &0.392 &0.481 & &AD~Pup &1.1334 &0.253 &0.810 & &X~Vul &0.8007 &0.792 &0.598 \\
SZ~Cas &1.1346 &0.961 &0.522 & &X~Lac &0.7360 &0.375 &0.525 & &AQ~Pup &1.4769 &0.516 &0.839 & &SV~Vul &1.6533 &0.489 &0.960 \\
TU~Cas &0.3303 &0.119 &0.497 & &Y~Lac &0.6359 &0.205 &0.526 & &BD~Pup &0.5931 &0.670 &0.565 & &HD~18391 &2.2500 &1.011 &0.955 \\
XY~Cas &0.6534 &0.518 &0.625 & &Z~Lac &1.0369 &0.411 &0.686 & &BE~Pup &0.4581 &0.640 &0.470 & & & \\
BD~Cas &0.5624 &1.017 &0.557 & &RR~Lac &0.8073 &0.365 &0.519 & &BN~Pup &1.1359 &0.457 &0.713 & & & \\
\hline
\end{tabular*}
\end{table*}

The periods used for the Cepheids in Table~\ref{tab4} are either the observed periods, if considered to originate from pulsation in the fundamental mode, or the observed or inferred fundamental mode period $P_0$ for double mode pulsators or Cepheids, such as QZ~Nor, SU~Cas, and V1334~Cyg, considered to be pulsating in the first overtone mode with period $P_1$. Decisions about pulsation mode may be incorrect in some instances.

\section{{\rm \footnotesize DISCUSSION}}
\begin{figure}[!t]
\begin{center}
\includegraphics[width=0.45\textwidth]{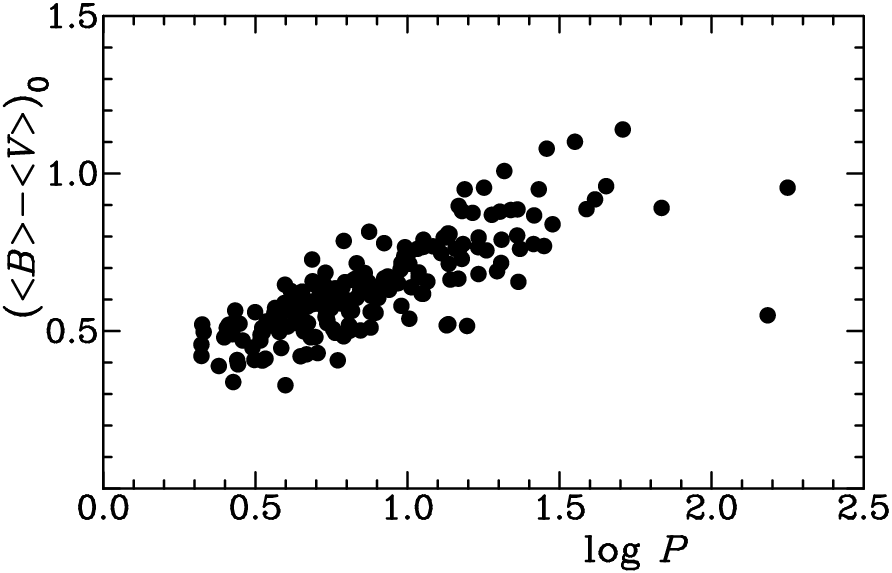}
\end{center}
\caption{\small{Derived intrinsic $(\langle B \rangle - \langle V \rangle)_0$ colors for the Cepheids in Table~\ref{tab4} plotted as a function of the logarithm of the pulsational period $P_0$.}}
\label{fig11}
\end{figure}

The original intent of the present study was to demonstrate that it is possible to generate a system of Cepheid color excesses that is solidly tied to the available sources of space reddening for a significant sample of stars, including in the compilation color excesses derived from reddening-free indices as a means of augmenting the new standard system (Std). An earlier version of such a compilation was described by \citet{tu01}, where it was noted that some generalaties that could be inferred from such a compilation regarding the Cepheid instability strip differed from those described by \citet{fe90b}. Similar conclusions can be made from the Cepheid sample of Table~\ref{tab4}. Fig.~\ref{fig11}, for example, is an empirical mapping of the colors of Cepheids as a function of pulsation period, as derived from the color excess summary, and is very similar to Fig.~2 of \citet{tu01}. Note, for example, that the color width of the instability strip changes very little as a function of pulsation period, contrary to the conclusions of \citet{fe90b}.

The methodology described by \citet{tb02} and \citet[][see \S1]{te10} can be used with the Table~\ref{tab4} data to map the instability strip in terms of effective temperature and luminosity, as done by \citet{tu12b}. Such results are not repeated here, although it is noted that they support the general conclusions made above. There is less scatter of Cepheids in the instability strip than is sometimes believed, although caution is needed when interpreting results for individual Cepheids. For example, the most deviant data in Fig.~\ref{fig11} all relate to stars of special interest. A few are likely Type II Cepheids, for example, and the two stars of longest period are V810~Cen and HD~18391, respectively. The former appears to be double-mode Cepheid-like variable, the latter a small-amplitude double-mode pulsator, both lying blueward of the Cepheid instability strip.

A much smaller reddening is cited for HD~18391 by \citet{ap90}, who tabulated, but did not use, a value of E$_{B-V} = 0.56$ by \citet{sc84} for the reddening of the double cluster $h$ and $\chi$ Persei. That value is clearly much too small according to the study of HD~18391's surroundings made by \citet{te09a}, and would lead to an intrinsic {\it B--V} color for the supergiant almost a half magnitude redder than expected for a G0~Ia star lying off the blue edge of the instability strip \citep[see][]{te09a}. Supergiants in the field around the double cluster also display strong evidence for differential reddening, with E$_{B-V}$ ranging from 0.21 to 1.00, or larger, in this section of the Perseus spiral arm. The reddening in the core regions of $h$ and $\chi$ Persei is therefore not representative of their outlying regions.

A useful extension of the present study would be to include light amplitude as an additional parameter with the Table~\ref{tab4} data plotted in Fig.~\ref{fig11}. Light amplitude has previously been demonstrated to be related to a Cepheid's location in the instability strip, as inferred from its rate of stellar evolution \citep[see][]{te06a}. An additional mapping of light amplitude within the strip using unreddened colors would help to confirm such findings. But that is left for a later investigation. The results of the present study are most usefully summarized by the compilations of Tables~\ref{tab2} and \ref{tab4}.

As noted earlier, several investigations of Cepheid field reddenings through membership in open clusters are as yet incomplete. It might also be possible to augment the sample of Cepheid reddenings by applying the transformation relations of Table~\ref{tab3} to other Cepheids in the cited studies that are not coincident with the main sample of this paper. A larger sample of semi-empirical reddenings for classical Cepheids is therefore within relatively easy reach.

\section*{{\rm \footnotesize ACKNOWLEDGEMENTS}}
\scriptsize{The present study owes much to the efforts of many different research collaborators over the author's lifetime, for which he is eternally grateful. The referee, Armando Arellano Ferro, also did much to inspire a thorough reworking of the original results to place them on sounder footing and make them more useful to a broader community. The study is much better for that input.}

\end{document}